\documentclass[number]{elsarticle} %,preprint elsarticle

\usepackage[english]{babel}
\usepackage[utf8]{inputenc}
\usepackage[T1]{fontenc}
\usepackage{ae,aecompl}

\usepackage{amsmath} 
\usepackage{amssymb}
\usepackage{commath}
\usepackage{dsfont}
\usepackage{bm}

\usepackage{graphicx}
\usepackage{subfigure}
\usepackage{booktabs}
\usepackage{natbib}
\usepackage{url}
\usepackage{color}
\usepackage{multicol}

\newcommand{\fanswer}[1]{\textit{\textcolor{red}{#1}}}
\newcommand{\sanswer}[1]{\textit{\textcolor{blue}{#1}}}

\newcommand{\numerico}{NUMERICO}

\newcommand{\numericoEN}{Nucleus of Modeling and Experimentation with Computers -- \numerico{}}

\newcommand{\R}{\ensuremath{\mathbb{R}}}            % real numbers
\renewcommand{\vec}[1]{\bm{#1}}

\DeclareMathOperator*{\argmin}{arg\,min}
\begin{document}

% --------------------------------------------------------------
\begin{frontmatter}

\journal{Applied Mathematics and Computation}

\title{Calibration of a SEIR--SEI epidemic model to describe\\ the Zika virus outbreak in Brazil}

\author[uerj]{Eber Dantas}
\ead{eber.paiva@uerj.br}

\author[uerj]{Michel Tosin}
\ead{michel.tosin@uerj.br}

\author[uerj]{Americo Cunha Jr\corref{cor}}
\ead{americo@ime.uerj.br}

\cortext[cor]{Corresponding author.}
%\cortext[a1]{ORCiD: https://orcid.org/0000-0003-2693-0719}
%\cortext[a2]{ORCiD: https://orcid.org/0000-0002-0112-553X}
%\cortext[a3]{ORCiD: https://orcid.org/0000-0002-8342-0363}

\address[uerj]{Universidade do Estado do Rio de Janeiro -- UERJ, \numericoEN{},
			  Rua S\~{a}o Francisco Xavier, 524, Rio de Janeiro - RJ, 20550-900, Brasil.
			  Phone: +55 21 2334-0323 r: 208
}

% Abstract
\begin{abstract}
Multiple instances of Zika virus epidemic have been reported 
around the world in the last two decades, turning the related illness into an international concern. 
In this context the use of mathematical models for epidemics is of great importance, since they 
are useful tools to study the underlying outbreak numbers and allow one to test the 
effectiveness of different strategies used to combat the associated diseases. This work deals with 
the development and calibration of an epidemic model to describe the 2016 outbreak 
of Zika virus in Brazil. A system of 8 differential equations with 8 parameters
is employed to model the evolution of the infection through two populations.
Nominal values for the model parameters are estimated from the literature.
An inverse problem is formulated and solved by comparing the system response to real
data from the outbreak. The calibrated results presents realistic parameters 
and returns reasonable descriptions, with the curve shape similar to the outbreak evolution 
and peak value close to the highest number of infected people during 2016. Considerations
about the lack of data for some initial conditions are also made through an analysis over
the response behavior according to their change in value.

\end{abstract}

% Keywords

\end{frontmatter}
% --------------------------------------------------------------

% --------------------------------------------------------------
\pagebreak
\section{Introduction}

The Zika virus is a flavivirus that upon infection in humans causes an illness, known as Zika fever,
identified commonly with macular or papular rash, mild fever and arthritis 
\cite{Brasil2016,WHOurl2016}. It is mainly a vector-borne disease carried by 
the genus \textit{Aedes} of mosquitoes \cite{WHOurl2016,Fernandes2016}, while in a lesser amount
it is also transmitted via sexual interaction \cite{Petersen2016,Coelho2016}, 
and contamination by blood transfusion is under investigation \cite{Motta2016}. 
The Zika virus was first isolated in primates from the Zika forest in Uganda in $1947$ \cite{Dick1952}. 
Evidences of the virus in humans were found in Nigeria in $1968$ \cite{Moore1975}. 
An epidemic occurred in $2007$ on Micronesia \cite{Duffy2009}, followed 
by multiple outbreaks on several Pacific Islands between $2013$ 
and $2014$ \cite{Cao2014,Tognarelli2016}. The first Zika virus autochthonous case 
in Brazil was reported around April, $2015$ \cite{SVS2017}, and nearly $30{,}000$ cases of infection 
were already notified by January $30$, $2016$ \cite{Faria2016}, 
along with the Pan American Health Organization being informed in the same month about 
locally-transmitted cases on numerous continental and island territories of America \cite{Hennessey2016}. 
The Brazilian Ministry of Health registered $215{,}319$ probable cases of Zika fever 
($130{,}701$ of which were confirmed) until the $52$th epidemiological 
week (EW) of $2016$ \cite{SVS2017}. The Zika epidemic has been causing concern in 
the international medical community, health authorities and population, specially due to an association 
between the Zika virus and other diseases, such as newborn microcephaly \cite{Petersen2016,Valentine2016} 
and  Guillain-Barré syndrome \cite{Santos2016}, whose correlation to the 
Zika virus was considered by the World Health Organization 
a \textit{``scientific consensus''} \cite{WHOApril2016}.

In this epidemic scenario, the development of control and prevention strategies for the mass 
infection is a critical issue. A mathematical model capable of \fanswer{providing} a description 
of the infected people throughout an outbreak is \fanswer{a useful} tool that can be employed to identify
effective and vulnerable aspects on disease control programs \cite{Naheed2014,Huppert2013,Lizarralde2017}.
Furthermore, for an epidemic model to be truly useful it must undergo a judicious process of \emph{validation}
\cite{Oberkampf2010,Cunhajr2017}, which consists in comparing model predictions with real 
data in order to evaluate if they are realistic. In general, the first predictions of a model 
do not agree with the observations, possibly due to inadequacies in the model hypotheses or 
because of a poor choice for the model nominal parameters. The first case invalidates the model, 
but the second can be amended through a procedure known as \emph{model calibration}, where a 
set of parameters that promote a good agreement between predictions and observations is sought. 

This work is one of the results in a rigorous ongoing process of identification and 
validation of representative models to describe Zika virus outbreaks in Brazil
\cite{Cunhajr_ccis2016,Cunhajr_cnmac2017_1}.
For this purpose, a SEIR-SEI mathematical model is adapted to the Brazilian scenario.
This specific SEIR-SEI description has been successfully used before for the outbreaks in
Micronesia \cite{Funk2016} and French Polynesia \cite{Kucharski2016}. Some
assumptions were also based on similar studies performed over SEIR dynamical systems \cite{Wang2014,Wanga2017,Safi2013,Cai2017}. The nominal values of the model parameters 
belong to characteristics of the Zika infection 
and its vector, quantitatively estimated in the literature or published by health 
organizations. Predictions are obtained from numerical simulation and further heuristic 
manipulation, followed by a comparison to real data of the outbreak as 
an initial effort to validate the model. In sequence, a rigorous process of model calibration 
is performed through the formulation and solution of an inverse problem.

The rest of this paper is organized as follows. In section~\ref{EpidemicModel}, 
the mathematical model is described and estimation of nominal values for
the model parameters is discussed. In section~\ref{model_calibration}, 
the forward and inverse problems are formulated and solved, where results are detailed
and a subsequent comparison between model predictions 
and experimental data is made to calibrate the model.
Finally, in section~\ref{concl_remaks}, the main contributions of this work 
are emphasized and a path for future works is suggested.
% --------------------------------------------------------------

% --------------------------------------------------------------
\section{Epidemic model for Zika virus dynamics}
\label{EpidemicModel}

\subsection{Model hypotheses}
\label{ModHyp}

\begin{figure}[h!]
\centering
\includegraphics[height=5.7cm]{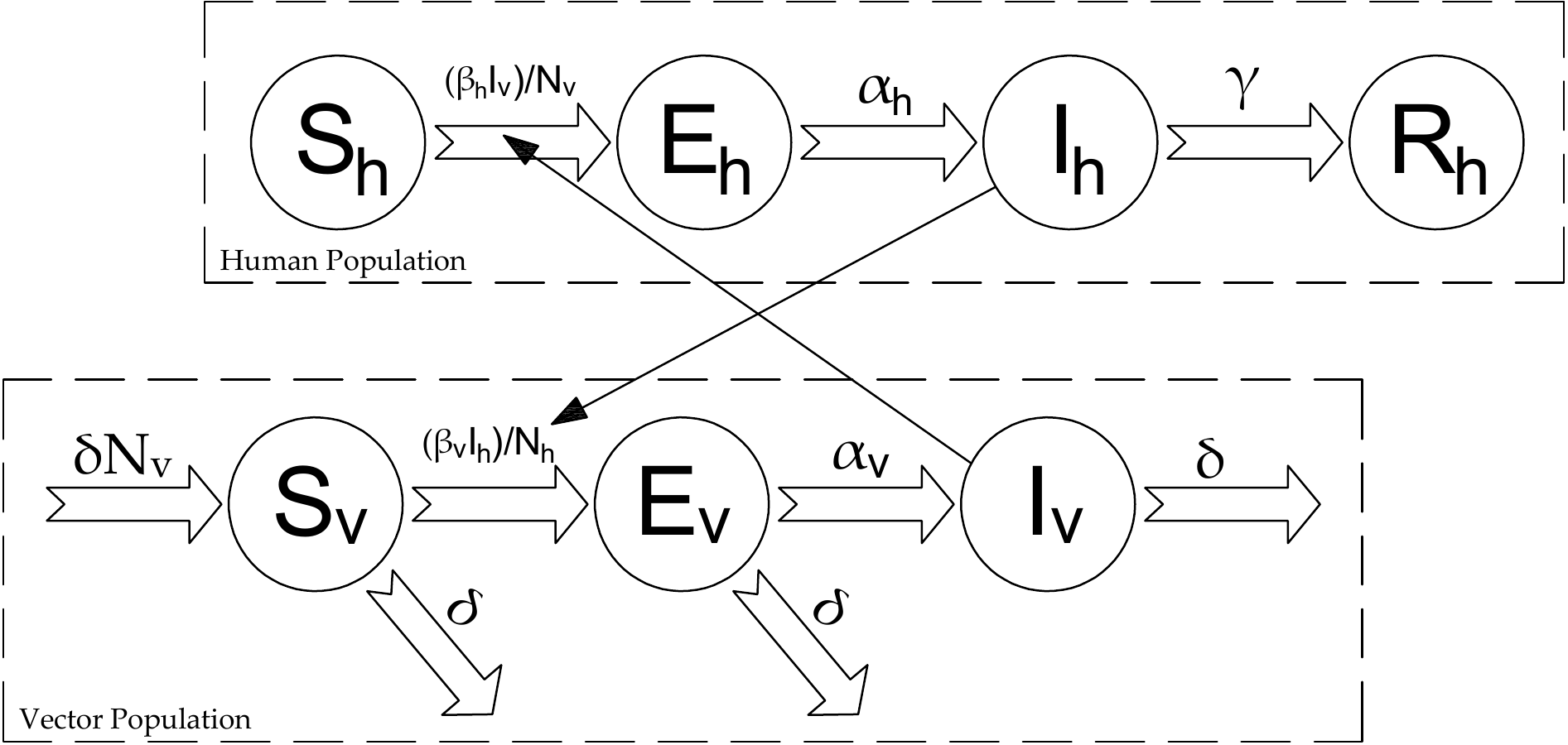}
\vspace{-2mm}
\caption{Schematic representation of the SEIR--SEI model for the Zika virus outbreak description.}
\label{model_schematic}
\end{figure}

This work utilizes a variant of the Ross-Macdonald model \cite{Smith2012} 
for epidemic predictions, separating the populations into a SEIR-SEI framework 
(susceptible, exposed, infectious, recovered) \cite{Brauer2012,Brauer2008,Martcheva2015}. 
Each category represents the health condition of an individual inside such group at time $t$, 
with respect to the considered infection. The susceptible group, denoted by $S(t)$, represents 
those who are uncontaminated and are able to become infected. The exposed portion of the population, 
$E(t)$, comprehends anyone that is carrying the pathogen but is still incapable of transmitting the disease. 
While the infectious individuals, $I(t)$, can spread the pathogen and may display symptoms associated 
with the illness. Finally, the recovered group, $R(t)$, contains those who are no longer infected. The 
populations of humans and vectors are segmented into the aforesaid classes (excepting the group of 
recovered vectors), as Figure~\ref{model_schematic} depicts 
schematically with the accordingly subscripts. $S_h$, $E_h$, $I_h$ and $R_h$ amass the number of 
people at each stage of the model description, and $S_v$, $E_v$, $I_v$ signifies proportion of vectors
($0 \leq S_v, \, E_v, \, I_v \leq 1$ and $ S_v + E_v + I_v = 1$).

\fanswer{Demographic} changes in the number of humans are not considered because the timescale of infection
is much faster than the timescale of birth and deaths for the latter to significantly alter the
development of the disease in the Brazilian context (supplementary material B provides tools to test this
assertion). The total vector population is maintained constant during the analysis, 
although variations of the proportion of vectors on the particular 
compartments are introduced via birth and death rates. The vector in 
question is regarded as a hypothetical mosquito apt to being infected 
or infectious throughout all its lifetime --- which means the model 
accounts only for the adult stage of their life cycle --- and also unable 
to recover.

The time elapsed while an individual is on the aforementioned 
exposed group is known as the latent period of an organism and, in this work, 
is adopted as equivalent to the commonly called incubation period (the time elapsed 
between being infected and exhibiting symptoms), since data is extremely 
sparse on the latter for humans \cite{Chan2012} (namely, the intrinsic incubation period). 
Both terms are used interchangeably hereafter, and the concepts do not differ 
on the mosquitoes case (the extrinsic incubation period) \cite{Lessler2016}. 
In addition, all the members of the susceptible group are treated as equally capable 
of being infected and the recovered ones as completely immunized.
% --------------------------------------------------------------

% --------------------------------------------------------------
\subsection{Model equations}
\label{ModEq}

The evolution of individuals through the SEIR-SEI groups is governed by the following 
(nonlinear) autonomous system of ordinary differential equations
\begin{equation}
\begin{aligned}
\dod{S_h}{t} &= - \beta_h \, S_h \, \frac{I_v}{N_v} \:, 
			& \dod{S_v}{t} &= \delta\,N_v - \beta_v \, S_v \, \frac{I_h}{N_h} - \delta \, S_v \:, \\[1em]
\dod{E_h}{t} &= {\beta_h} \, S_h \, \frac{I_v}{N_v} - \alpha_h \, E_h \:,
			& \dod{E_v}{t} &= \beta_v \, S_v \, \frac{I_h}{N_h} -(\alpha_v + \delta) \, E_v \:,\\[1em] 
\dod{I_h}{t} &= \alpha_h \, E_h - \gamma \, I_h \:,
			& \dod{I_v}{t} &= \alpha_v \, E_v - \delta \, I_v \:, \\[1em]
\dod{R_h}{t} &= \gamma \, I_h \:, 
			& \dod{C}{t} &= \alpha_h \, E_h \:,
\end{aligned}
\label{model_eqs}
\end{equation}
\noindent
where $N$ represents the total population and $1/\!\alpha$ the disease's incubation period 
(each with the corresponding subscript of $h$ for human's and $v$ for vector's), 
$1/\!\delta$ means the vector lifespan, $1/\!\gamma$ is the human infection period --- which is defined 
in this work as the interval of time that a human is infectious --- and $\beta$ identifies the transmission rate, 
specifically $\beta_h$ is the mosquito-to-human rate and $\beta_v$ the human-to-mosquito rate.

The transmission terms $\beta_h\,S_h\,I_v/N_v$ and $\beta_v\,S_v\,I_h/N_h$ are composed by a 
number of susceptible individuals ($S_h,\,S_v$, respectively), a transmission rate 
($\beta_h,\,\beta_v$) and the probability of the contact being made with an infectious member 
of the other population ($I_v/N_v,\,I_h/N_h$). Both transmissions terms come from the assumption 
that the rate of contacts is constant, which characterizes a frequency-dependent transmission
\cite{Begon2002}.

The compartmentalization hypothesis requires setting the variables at the initial time of the
analysis, $t_0$, such that their sum equals the total population in each case, e.g, $S_v(t_0)+E_v(t_0)+
I_v(t_0) = N_v = 1$. \fanswer{By so doing}, for all $t>t_0$, $S_h(t)$, $E_h(t)$, $I_h(t)$ and $R_h(t)$ will 
always add up to $N_h$; and $N_v(t)$ will be the equilibrium solution of the initial value problem
\begin{equation}
\begin{aligned}
\dod{N_v}{t} &= \delta\,(1 - N_v(t)) \:, & N_v(t_0) = 1 \:,
\end{aligned}
\end{equation}
\noindent
namely, $N_v(t) = 1$. This consideration allows the simplification of treating the total vector population
as a parameter $N_v$, instead of a variable, since it stays constant throughout the analysis. 
If one wishes to treat the vector population as a variable, a recruitment parameter would need to 
be added in place of $\delta\, N_v$ as well as another differential equation to account for the changes in
$N_v(t)$.

The $\dif C/\!\dif t$ equation allows evaluation of the cumulative number $C(t)$ of infectious people
until the time $t$, that is, the amount of humans so far that contracted the disease and have 
passed through or are in the infectious group at the given time.

Additionally, a set of $M=52$ points to represent the number of new infectious cases of Zika fever
at each week is defined as follows:
\begin{equation}
\mathcal{N}_{w} = C_{w} - C_{w-1} \:, \quad \mathcal{N}_{1} = C_1, ~~w=2, \cdots, 52 \:,
\label{NC_eq}
\end{equation}
\noindent
where $C_{w}$ is the cumulative number of infectious humans 
in the \emph{w}-th EW.

\begin{figure}[h]
\centering
\subfigure[Cumulative number]{\includegraphics[scale=0.32]{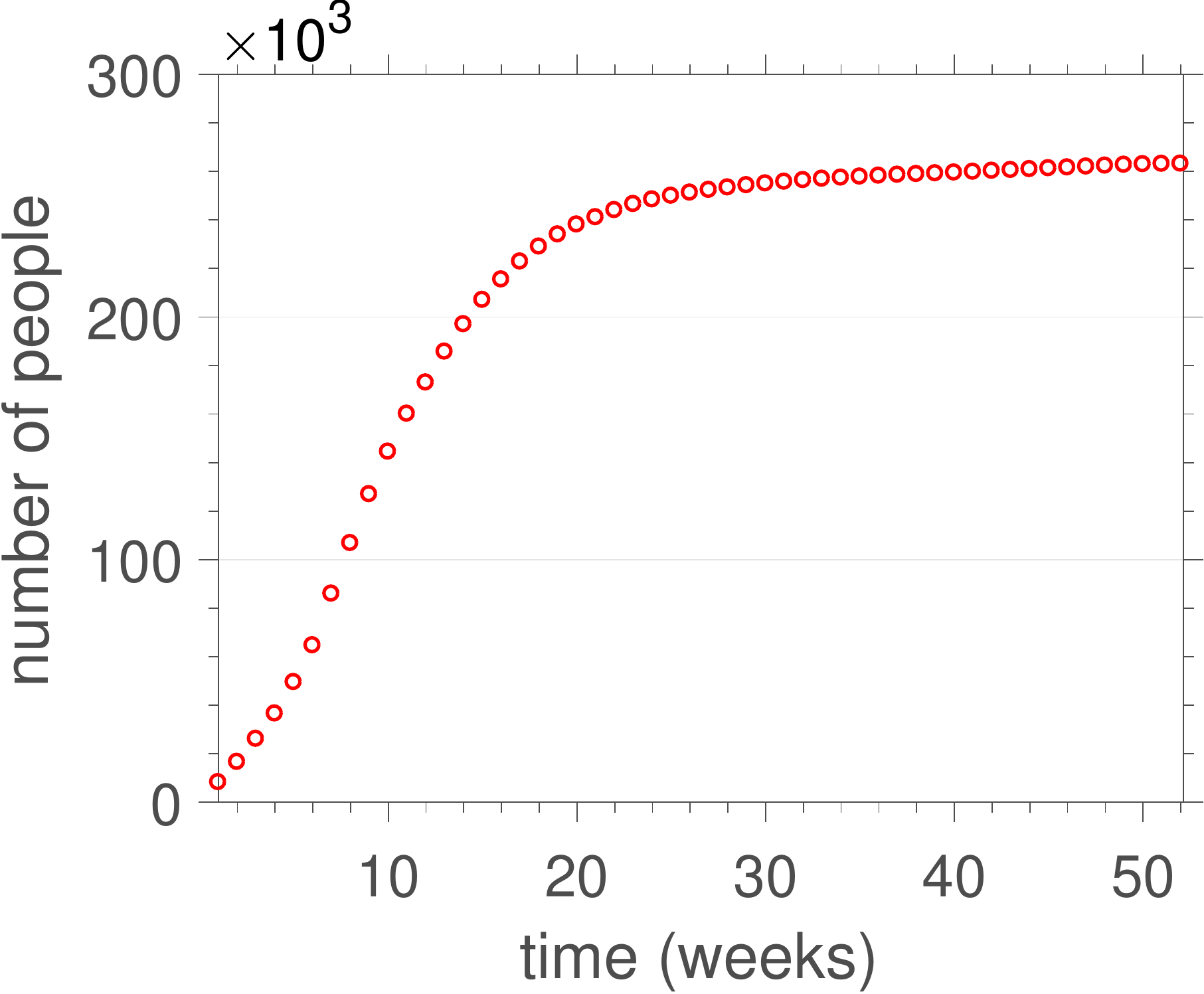}}
\subfigure[New cases]{\includegraphics[scale=0.32]{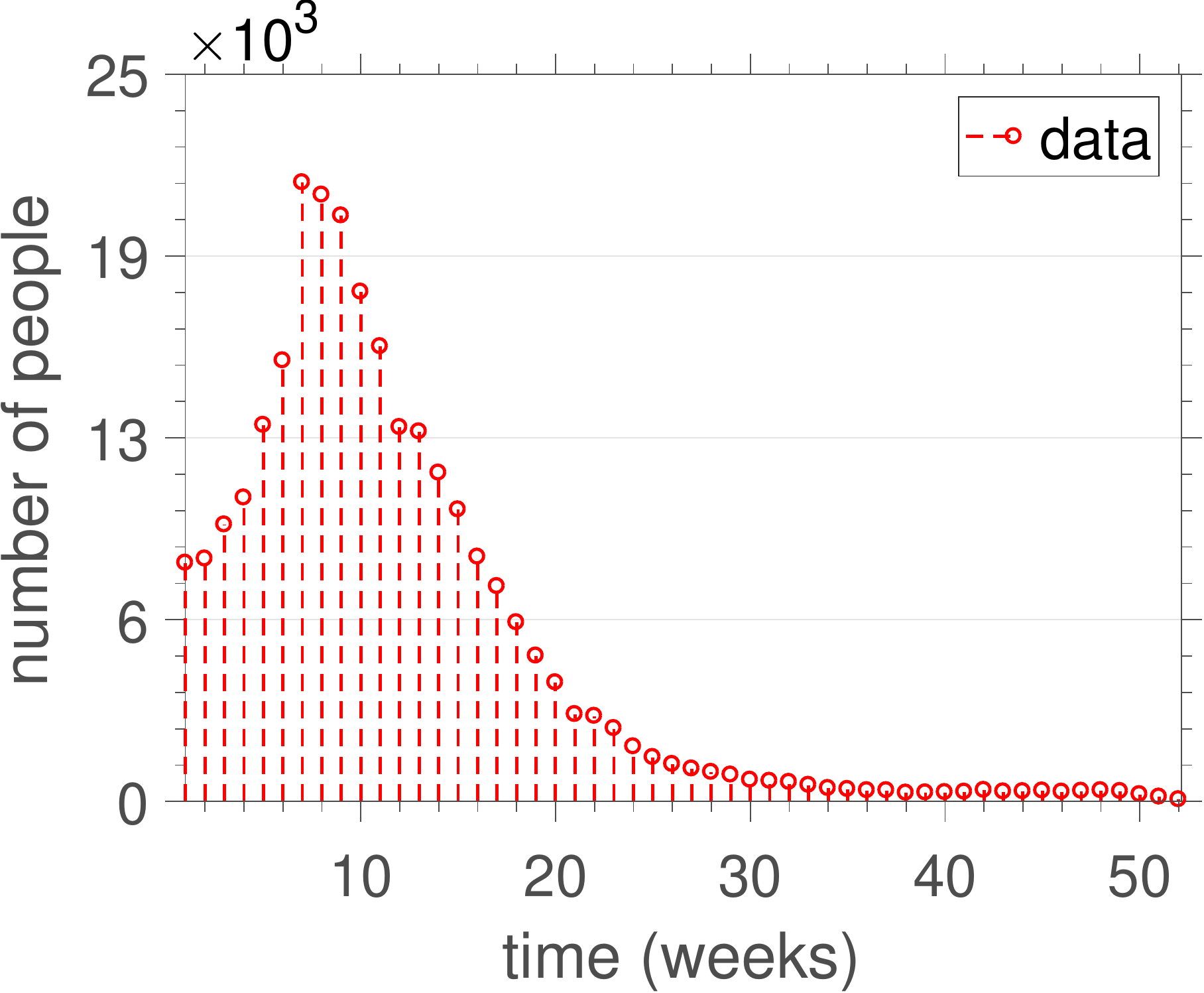}}\\
\vspace{-3mm}
\caption{Outbreak data.}
\label{data}
\end{figure}

Figure~\ref{data} organizes the data of cumulative number of infectious and new cases per 
week provided by the Brazilian Ministry of Health \cite{Data2016} (supplementary material A) for 2016, 
where \sanswer{the evolution of the infection can be seen}.
The expected behavior for the model is that it generates a seemingly year long outbreak for 2016
that matches $C(t)$ and $\mathcal{N}_{w}$ to the data of Figure~\ref{data}. The end of this outbreak is 
marked by the exposed and infectious portions of vectors reaching very low values, effectively killing off
the contamination process to the point that $S_h(t)$ stabilizes.

% --------------------------------------------------------------

% --------------------------------------------------------------
\subsection{Nominal parameters}
\label{NomParam}

The preliminary values for the parameters of the set of Eqs.~(\ref{model_eqs}) come from the 
related literature concerning the Zika infection,  vector-borne epidemic models, 
the \textit{Aedes aegypti} mosquito (which is the main vector for 
Zika, Dengue and Yellow fever in Brazil) and publications provided by 
health organizations and government agencies. 
The Brazilian Institute of Geography and Statistics 
reports that Brazil had approximately $N_h = 206 \times 10^6$ people by July, $2016$ \cite{DOU},
and $N_v$ is set $1$ to entail an entire vector population.
The adopted extrinsic incubation period is  $1/\!\alpha_v = 9.1$ days \cite{Funk2016}.
This value agrees with other statistical confidence intervals (CI) 
that are presented for the parameter in another works ($95$\% CI: $7.3$--$9.3$ days \cite{Ferguson2016})
and is close to the numbers suggested by experimental studies for the time necessary for the virus 
to reach the mosquito's saliva after an infectious blood meal ($5$ \cite{Wong2013,Li2012} 
and $7$ days \cite{Chouin-Carneiro2016}). 
A systematic review and pooled analysis of the literature and 
case studies available in \cite{Ott2016} estimates that the median intrinsic incubation period is
$5.9$ days ($95$\% CI: $4.4$–-$7.6$). This values is selected for $1/\!\alpha_h$ in this work and is
compatible with the range of $3$--$12$ days recommended by multiple sources
\cite{Valentine2016,Ioos2014,ECDC2015},
also formerly used in previous studies \cite{Villela2017}.
The aforementioned literature 
analysis in \cite{Ott2016} also concludes that $9.9$ ($95$\% CI: $6.9$-–$21.4$) days is the mean time 
until an infected has no detectable virus in blood. Considering the assumption that the infectiousness 
in Zika infection ends $1.5$--$2$ days before the virus becomes 
undetectable \cite{Funk2016,Ferguson2016}, the chosen value for the human infectious 
period is $1/\!\gamma = 9.9-2 = 7.9$ days.
As for the vector lifespan $1/\!\delta$, 
``\textit{the adult stage of the mosquito is considered to last an average of eleven days in 
the urban environment}" according to \cite{Otero2006}. This is the assumed value for the parameter
in this work, which is also consistent with the usual life expectancy for 
the mosquito in Rio de Janeiro, Brazil
\cite{Freitas2007}, and comes close to the average of $2$--$3$ weeks considered in
biological studies about the species \cite{Nelson1986} and by the Centers for Disease Control and
Prevention \cite{CDCurl}. 
Lastly, the time between a 
mosquito being infected and it infecting a human, $1/\!\beta_h$, and the time between a human 
infection and a mosquito taking an infectious blood meal, $1/\!\beta_v$, is estimated 
in \cite{Ferguson2016} as an average of $11.3$ days ($95$\% CI: $8.0$--$16.3$) 
and $8.6$ days ($95$\% CI: $6.2$-$11.6$), respectively.
% --------------------------------------------------------------

% --------------------------------------------------------------
\section{Calibration of the epidemic model}
\label{model_calibration}

\subsection{Forward Problem}
\label{forward_problem}

The epidemic model of section~\ref{EpidemicModel}, supplemented by an 
appropriate set of initial conditions, is a continuous-time dynamical system
\begin{equation}
	\dot{\mathbf{x}}(t) = \mathbf{f} \left( \mathbf{x}(t), \mathbf{p}\right) \:, \quad
	\mathbf{x}(t_0) = \mathbf{x}_{0}
	\label{ivp_def}
\end{equation}
\noindent
where $\mathbf{x}(t) = \left( S_h(t), E_h(t), I_h(t), R_h(t), S_v(t), E_v(t), I_v(t), C(t) \right) \in \R^8$ 
is the vector of states at time $t$, 
$\mathbf{x}_{0} = \left( S_h^i, E_h^i, I_h^i, R_h^i, S_v^i, E_v^i, I_v^i, C^i \right) \in \R^8$ 
is a prescribed initial condition vector referring to the initial time $t_0$ of
the analysis, the vector 
$\mathbf{p} = \left( N_h, \beta_h, \alpha_h, \gamma, N_v, \beta_v, \alpha_v, \delta \right) \in \R^8$
lumps the model parameters and 
$\mathbf{f}: U \subset \R^8 \times \R^8 \to \R^8$ is a nonlinear map which 
gives the evolution law of this dynamics, defined (for fixed $t$) on the open set
\begin{equation}
	U = \left\lbrace (\mathbf{x}(t),\mathbf{p}) \in \R^8 \times \R^8 ~\big |~ x_n(t) > 0 ~ \mbox{and} ~ p_n > 0, ~\mbox{for} ~ n=1,\cdots, 8 \right\rbrace .
\end{equation}

The forward problem consists in providing initial conditions (IC) and a set of parameters,
represented by the pair $\boldsymbol \alpha = (\mathbf{x}_{0},\mathbf{p})$,
and compute by means of numerical integration 
the model response $\mathbf{x}(t)$ from which a scalar 
observable $\boldsymbol \phi (\boldsymbol \alpha, t)$ is obtained. In the forward problem,
$\boldsymbol \alpha$ represents all IC and system parameters 
from Eq.(\ref{model_eqs}), while $\boldsymbol \phi (\boldsymbol \alpha, t)$ is the new cases 
$\mathcal{N}_{w}$ system response from Eq.(\ref{NC_eq}).

Since the map $\mathbf{f}$ has a polynomial nature in $\mathbf{x}$, it is a continuously 
differentiable function in $\mathbf{x}$. Thus, the existence and uniqueness theorem for ordinary
differential equations guarantees that the initial value problem of (\ref{ivp_def}) has an 
unique solution. Besides that, one can also show that this solution \sanswer{is} continuously dependent on
$\boldsymbol \alpha$, as well as the \emph{forward map} $\boldsymbol \phi$ \cite{Hirsch2013,Perko2006}.

The evaluation of the system response in the \emph{forward problem} is performed
numerically in this work via a Runge-Kutta (4,5) method and the scalar observable
of interest $\mathcal{N}_{w}$ is  used to assess the simulation when compared with real 
data of the 2016 outbreak made available by the Brazilian Ministry of Health 
\cite{Data2016} (supplementary material A). The referred data consists of probable cases of infected people per EW,
registered by sanitary outposts and health institutions throughout the country 
when the common symptoms of Zika fever were exhibited by an individual. 
In accordance with the hypothesis that one only displays symptoms when inside
$I_h(t)$, the $C(t)$ variable models this discrete accumulating data on a continuous sense and
$\mathcal{N}_{w}$ provides the corresponding influx per EW. The $C(t)$ time series 
is also observed as a criteria for good fitting, since it signifies the general impact
of the epidemic and because a reasonable $\mathcal{N}_{w}$ result does not necessarily implies 
an acceptable cumulative number of infectious individuals for all $t$ in comparison to the data.

The initial time of the analysis was established as the first EW of 2016. 
The remaining individuals in both populations are assumed susceptible at first, meaning 
$S_{h}^{i} = N_h - E_{h}^{i} - I_{h}^{i} - R_{h}^{i}$ and $S_{v}^{i} = N_v - E_{v}^{i} - I_{v}^{i}$. 
The initial values for the exposed and infectious groups are set equal, 
$E_{h}^{i} = I_{h}^{i}$ and $E_{v}^{i} = I_{v}^{i}$. 
Likewise, the number of infected humans at the initial time must be $I_{h}^{i} = C^{i}$, 
given its definition. The value of $C^i$ is taken as the number of confirmed Zika cases in Brazil 
on the first EW of $2016$ \cite{Data2016}, $8{,}201$ individuals, and the recovered
 group is assumed equal to the suspected number of infected in $2015$, according
to the data available \cite{Faria2016}, $R_{h}^{i} = 29{,}639$ individuals. 
As for the proportion of infectious vectors in the first week, to work around the lack of 
data for this initial condition, repetitive manual 
estimations were tried until the resulted time series of $\mathcal{N}_w$ presented reasonable 
values compared to the real data.
It became clear that the system response is very
sensible to $I_{v}^{i}$, as slight variations in 
its value are required to achieve feasible results. In the process of choosing its value, 
the matching of the $\mathcal{N}_w$ curve's peak to the amplitude of infection is also a priority, 
since this is the main interest region for evaluation of the outbreak.
The nominal values of the parameters exhibited viable $\mathcal{N}_w$ curves around 
$I_{v}^{i}= 2.2 \times 10^{-4}$. 

\begin{figure}[h]
\centering
\subfigure[Cumulative number]{\includegraphics[scale=0.32]{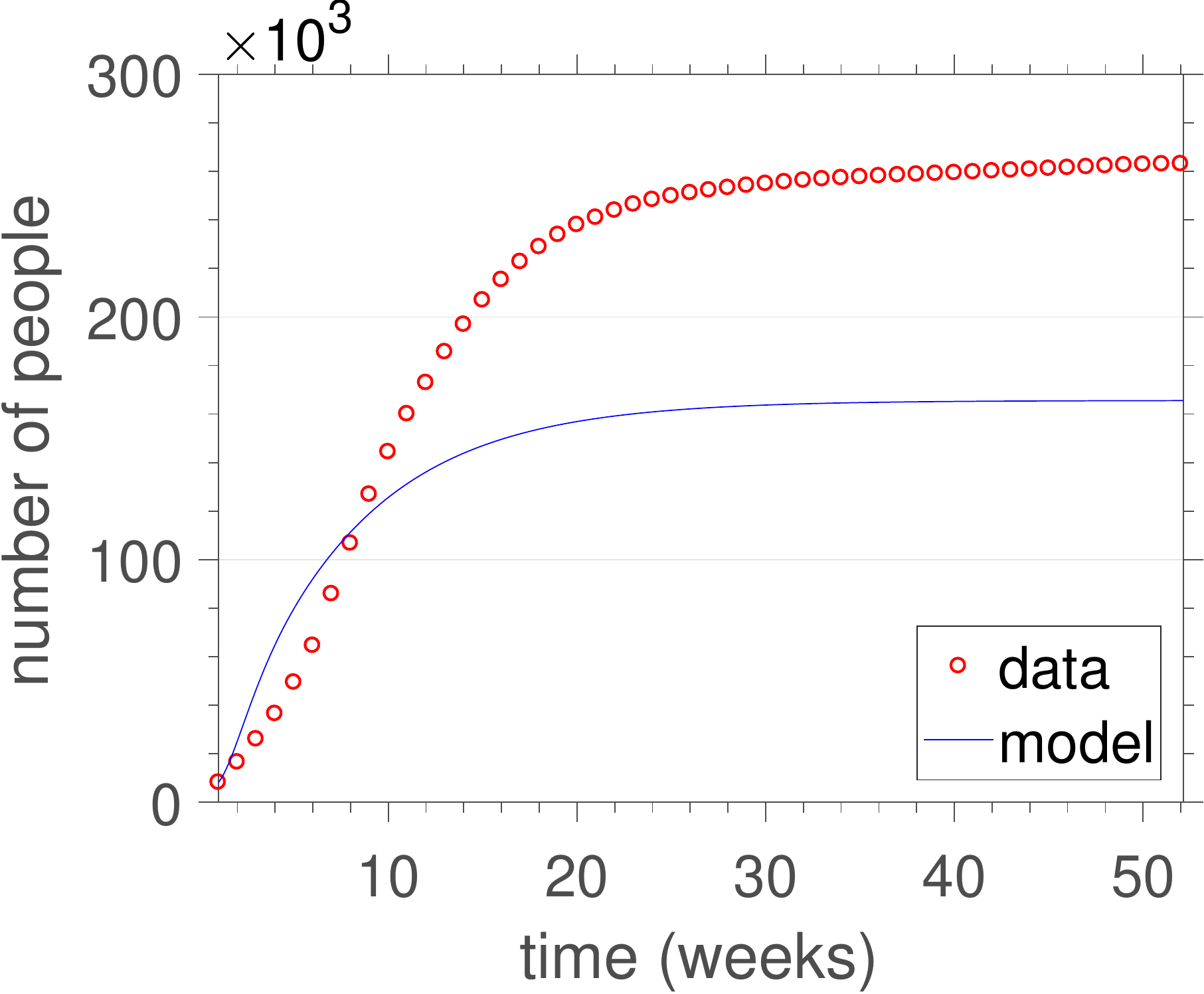} \label{graph_forward_C}}
\subfigure[New cases]{\includegraphics[scale=0.32]{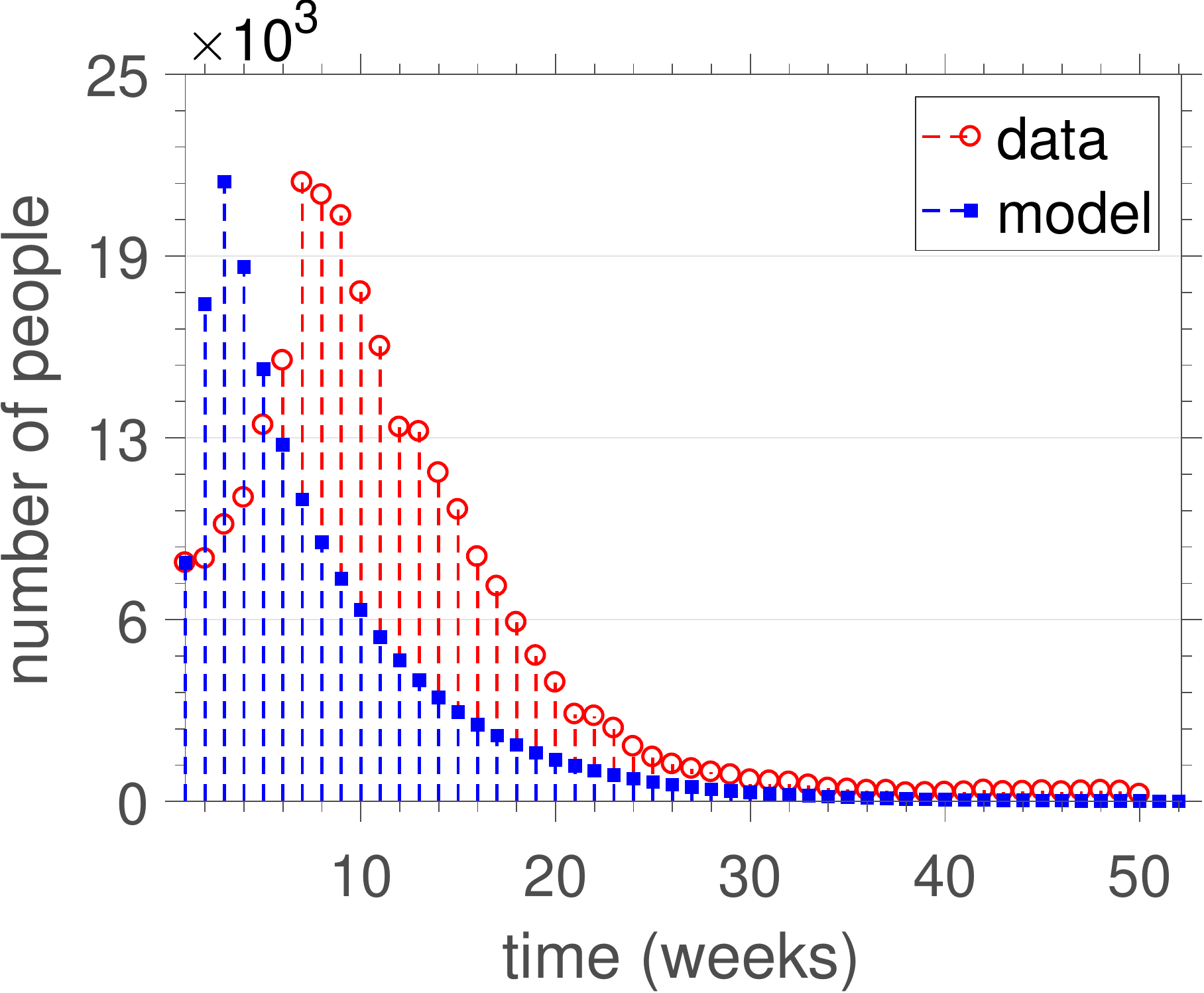}\label{graph_forward_NewCases}}\\
\vspace{-3mm}
\caption{Outbreak data (red) and model response using the nominal parameters (blue).}
\label{graph_forward_problem}
\end{figure}

Figure~\ref{graph_forward_problem} presents the SEIR-SEI model response for the nominal set of parameters
from section~\ref{NomParam}, supplied with the above IC, on \sanswer{an}
epidemiological week temporal domain consisting of one to fifty-two weeks ($7$ to $365$ days), compared 
with the data of the outbreak (red dots).

The general shape of the curves in Figure~\ref{graph_forward_problem} do provide qualitative information 
regarding the evolution of the infection, even though the portrayed descriptions
are not quantitatively realistic. This inherent pattern agreement and numerical 
mismatch \sanswer{suggests} that the model response may differ from the real data due to the use of 
unsuitable values for the parameters or incorrect IC assumptions.
The search for parameter and initial condition values that make the simulations fit well to 
the observed data defines model calibration (or system identification), 
being the object of interest of the next section.
%% ---------------------------------------------------------------

% --------------------------------------------------------------
\subsection{Inverse Problem}
\label{inverse_problem}

The model calibration problem seeks to find a set of parameters 
that, to a certain degree, makes the model response as close 
as possible to the empirical observations (reference data), once,
due to the erros on model conception and reference data acquisition, 
it is (practically) impossible for the  forward map to reproduce the 
outbreak observations.

The mathematical setting for this case considers the 
\emph{parameters vector} $\boldsymbol \alpha$ defined
in the \emph{parameter space} $E = \R^{12}$, since here
$\boldsymbol \alpha$ comprises all IC and system parameters 
from Eq.(\ref{model_eqs}) , excepting $N_h$, $N_v$, $R_h^i$ and $C^i$, 
which are kept fixed in their nominal values.
For the purpose of comparison between observations and predictions,
a discrete set with $M$ time-instants is considered, so that
scalar observations and predictions are respectively lumped into
$\vec{y} = (y_1, y_2, \cdots, y_M)$ and 
$\bm{\phi} \left( \bm{\alpha} \right) = \left(\phi_1, \phi_2, \cdots , \phi_M \right)$, 
both defined in the \emph{data space} $F=\R^{M}$. 
Note that the \emph{forward map} $\boldsymbol \phi: E \to F$ associates to
each \sanswer{parameters} vector $\bm{\alpha}$ an observable vector 
$\bm{\phi} \left( \bm{\alpha} \right)$ where the component
represents the number of new cases in each week of the year, 
i.e., $\bm{\phi}_{w} = \mathcal{N}_{w}$.
In practice, the parameters vector is restricted to be on the convex
\emph{set of admissible values} 
$C = \left\lbrace \boldsymbol \alpha \in E ~ \big | ~ \textbf{lb}\, \leq 
\boldsymbol \alpha \leq \,\textbf{ub} \right\rbrace$, in which
$\textbf{lb}$ and $\textbf{ub}$ are lower and upper vector bounds
for $\boldsymbol \alpha$, respectively.

In formal terms, given an \emph{observation vector} $\vec{y} \in F$ and a 
\emph{prediction vector} $\bm{\phi} \left(\boldsymbol \alpha \right) \in F$,
the calibration aims at finding a vector of parameters $\boldsymbol \alpha^{\ast}$ such that
\begin{equation}
\boldsymbol \alpha^{\ast} = \argmin_{\boldsymbol \alpha \in C} \, J(\boldsymbol \alpha)  \:, 
\label{inv_prob_def}
\end{equation}
\noindent
for a misfit function
\begin{equation}
J(\boldsymbol \alpha) = ||\vec{y} - \bm{\phi} \left(\boldsymbol \alpha \right) ||^2
= \left\lbrace \sum_{m=1}^{M} \Big| y_m - \phi_m \left(\boldsymbol \alpha  \right) \Big|^{2} \right \rbrace.
\label{misfit_func}
\end{equation}

\begin{figure}[h!]
\centering
\includegraphics[scale=0.40]{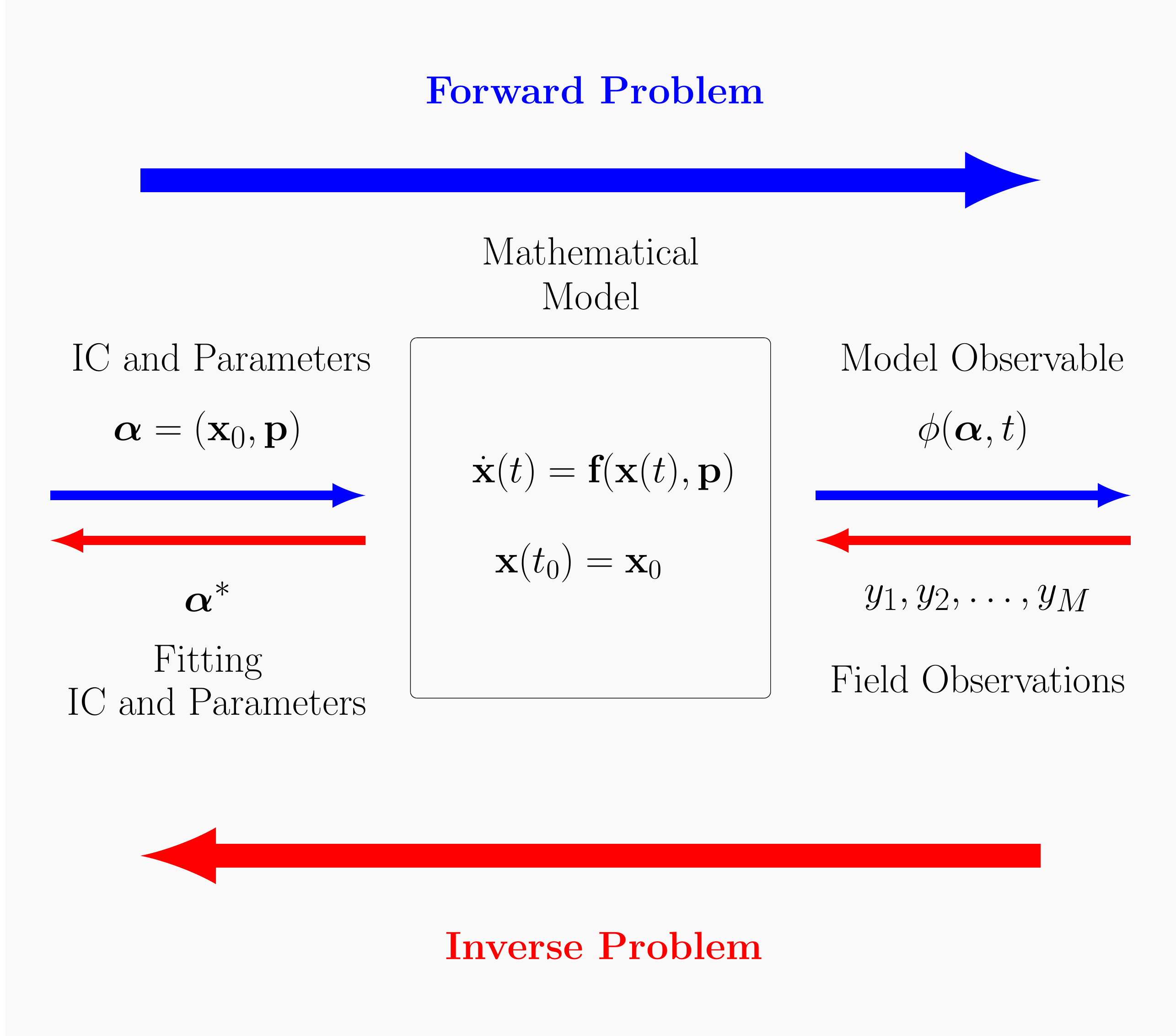}
\vspace{-3mm}
\caption{Schematic representation of the forward and inverse problems associated to the epidemic model.}
\label{InvProb_schematic}
\end{figure}

This is the \emph{inverse problem} associated 
to the epidemic model. In general this type of problem 
is extremely nonlinear, with none or low regularity, 
multiple solutions (or even none), 
being much more complicated to attack in comparison
with the forward problem
\cite{Aster2012,Yaman2013}.
A schematic representation of the forward and the inverse problem
associated to the epidemic model is shown in Figure~\ref{InvProb_schematic}.

This inverse problem attempts to estimate a finite number of parameters on a 
finite dimension space, being defined in terms of a typical nonlinear misfit function. 
\fanswer{Therefore, Theorem 4.5.1} of \cite{Chavent2010} can be invoked to guarantee 
a proper sense of well-posedness (existence, uniqueness, unimodality and local stability)
for the \fanswer{inverse} problem.

The Trust-Region-Reflective method \fanswer{(TRR)} is employed here to numerically approximate a 
solution for the inverse problem (\ref{inv_prob_def}). The main idea of the method is to minimize a simpler function 
that reflects the behavior of $J(\boldsymbol \alpha)$
in a neighborhood (trust-region) around $\boldsymbol \alpha$. The simpler function is defined as dependent on 
the trial step $\mathbf{s}$, characterizing the Trust-Region subproblem, and its computation is optimized 
by restricting the subproblem to a two-dimensional subspace. 
The subspace is linear spanned by a multiple of the gradient $\mathbf{g}$ and 
(in the bounded case) a vector obtained in a scaled modified Newton step, used for the 
convergence condition ${D(\boldsymbol \alpha)}^{-2}\,\mathbf{g}(\boldsymbol \alpha) = 0$,
where $D$ is a diagonal matrix that depends on $\boldsymbol \alpha$, $\mathbf{g}$, $\mathbf{lb}$, 
and $\mathbf{ub}$ \cite{Coleman1996}. Finally, the trial step is found through the subproblem as
\begin{equation}
\mathbf{s^*} = 
\argmin_\mathbf{s} \left\lbrace 
\frac{1}{2} \mathbf{s}^\mathrm{\!T} Q \mathbf{s} + \mathbf{g}^\mathrm{\!T} \mathbf{s} \: \mid \:
{||D\,\mathbf{s}||}_2 \leq \Delta \right\rbrace \:,  
\label{subproblem_eq}
\end{equation}
where $\Delta$ is a scalar associated with the trust region size; $Q$ is a matrix involving 
$D$, a Jacobian matrix (also dependent on $\boldsymbol \alpha$, $\mathbf{g}$, $\mathbf{lb}$, 
and $\mathbf{ub}$) and an approximation of the Hessian matrix \cite{Coleman1996}.
The quadratic approximation in Eq.(\ref{subproblem_eq}) has well-behaved solutions \cite{Conn2000}
and if $J(\boldsymbol \alpha + \mathbf{s}) < J(\boldsymbol \alpha) $
then $\boldsymbol \alpha$ is updated to $ \boldsymbol \alpha + \mathbf{s} $ and the process iterates, 
otherwise $\Delta$ is decreased. In addition, a reflection step also occurs if a given step 
intersects a bound: the reflected step is equivalent to the original step except in the 
intersecting dimension, where it assumes the opposite value after reflection. 

The TRR algorithm also requires an initial guess for each parameter, identified in the 
next section as ``TRR input''. The stopping criteria are the norm of the step and the 
change in the value of the objective function, with a tolerance of $10^{-7}$. 
Supplementary material B provides details on the \fanswer{software} used for implementation.
% --------------------------------------------------------------

% --------------------------------------------------------------

\subsection{Numerical experiments for calibration}
\label{NumExpCalib}

\begin{figure}[h]
\centering
\subfigure[Cumulative number]{\includegraphics[scale=0.32]{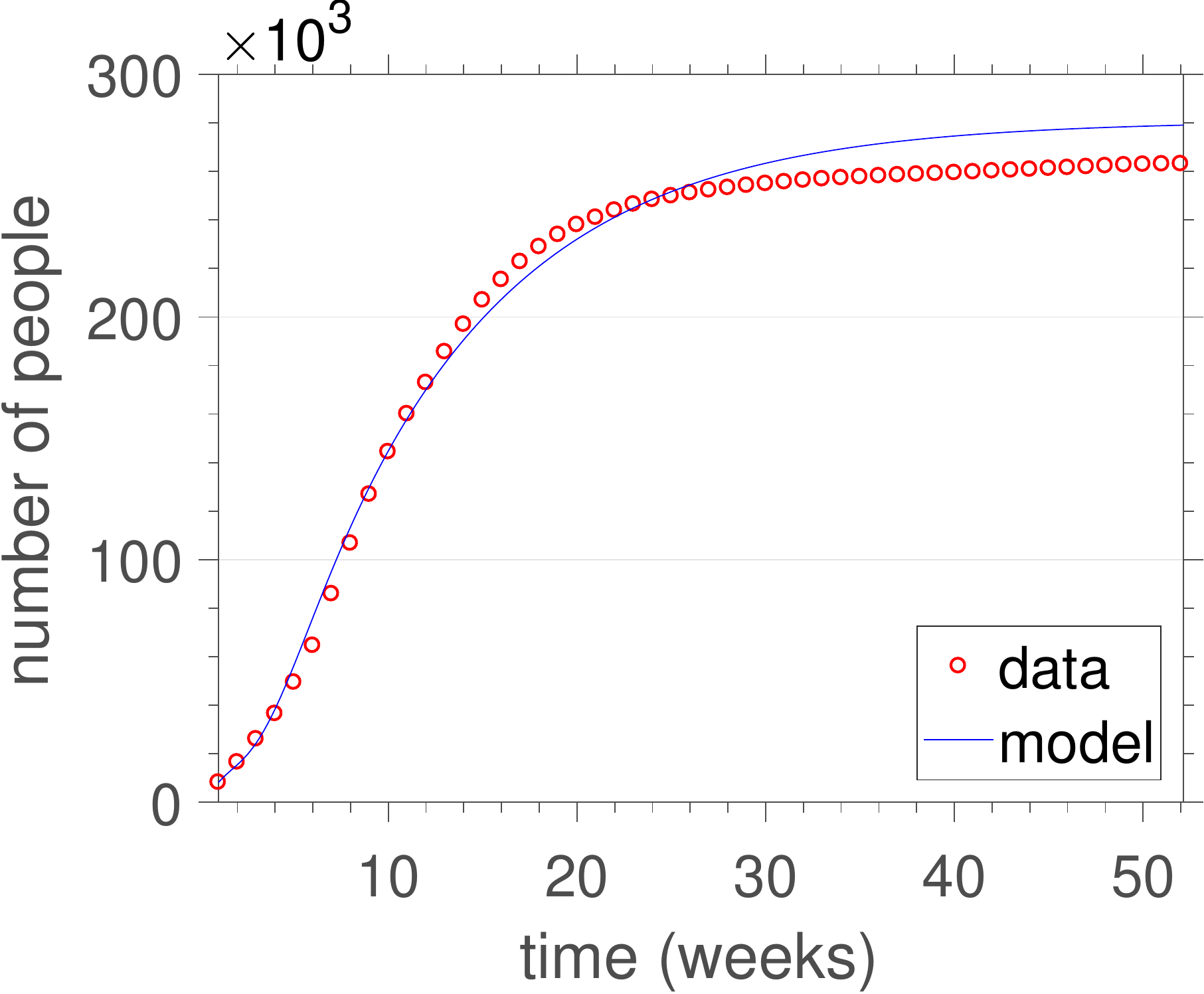} \label{graph_inverse_C_N}}
\subfigure[New cases]{\includegraphics[scale=0.32]{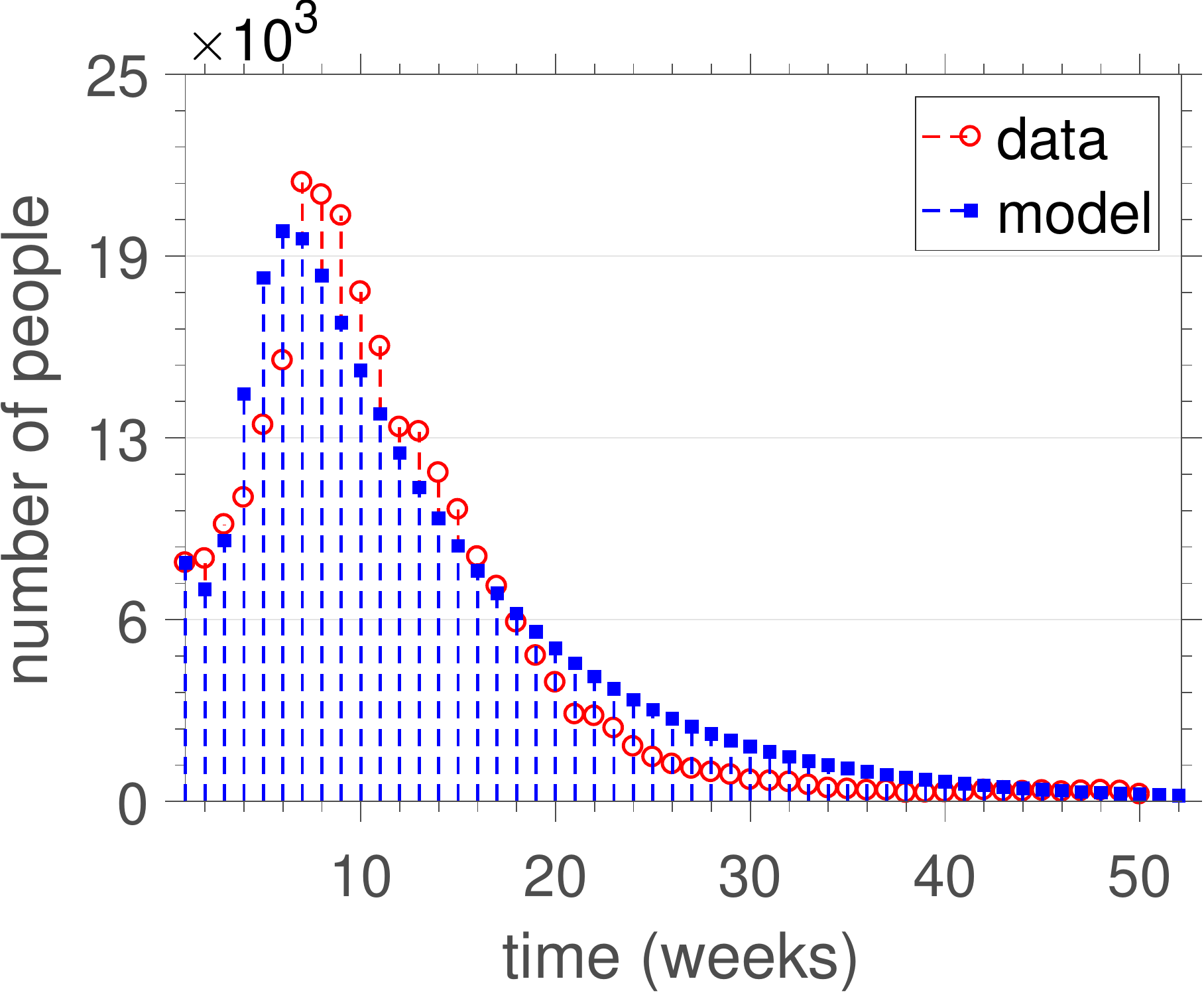}\label{graph_inverse_NewCases_N}}\\
\vspace{-3mm}
\caption{Outbreak data (red) and calibrated model response (blue) from Table~\ref{tab_inverse_N}.}
\label{graph_inverse_problem_N}
\end{figure}

Figure~\ref{graph_inverse_problem_N} presents the best result for the $\mathcal{N}_{w}$ system response 
fitting problem using the nominal parameters and IC from sections~\ref{NomParam} and \ref{forward_problem}
as initial guesses for the TRR algorithm. The upper and lower bounds used for the parameters
were set compatible with the literature suggested intervals and are presented in Table~\ref{tab_inverse_N},
along with the parameters and IC values resulted from the calibration (``TRR output''). The 
\sanswer{$\mathbf{ub}$} for $\delta$ was assumed \fanswer{lower} than the \sanswer{$\mathbf{lb}$} 
for $\alpha_v$ to maintain consistency with the model 
interpretation. The minima for $S_h^i$ and $S_v^i$ were set to 
$0.9\,N_h$ and $0.99$, respectively, to establish a high number 
of susceptible individuals as is expected for the beginning of an outbreak. Also, the 
\sanswer{lower and upper bound} ($0.999$) for $S_v^i$ were 
motivated by noticing how variations in $E_v^i$ and $I_v^i$ of order $10^{-3}$ already 
bring significant changes in the system response. The lack of available data 
for the exposed and infectious groups at the onset of the epidemic was circumvented by 
appointing its minimum and maximum possible values as $\mathbf{lb}$ and \fanswer{$\mathbf{ub}$}, i.e., 
$E_v^i$ and $I_v^i$ were restricted between zero and one, while
$E_h^i$ and $I_h^i$ were bounded by zero and $N_h$. 

Additionally, to ensure the model hypotheses of compartmentalization
and constant population, two additional fitting points were defined, $ \Sigma_h = S_h^i + E_h^i + I_h^i$ and 
$\Sigma_v = S_v^i + E_v^i + I_v^i$, which were set to match $N_h - R_h^i$ and
$N_v=1$ on Eq.(\ref{inv_prob_def}), respectively. However, the algorithm is only capable 
of \textbf{approximating} $\Sigma_h$ and $\Sigma_v$ to their intended values. So to 
account for these minor differences, the resulting values of  $(N_h - R_h^i) - \Sigma_h$ and 
$1 - \Sigma_v$ were added to the TRR output of $S_h^i$ and $S_v^i$. These corrections did not 
impact the calibration, since the scale of the differences would always be, correspondingly, 
$10^{-4}$ and $10^{-2}$ (at most), which are below the sensibility of  $S_h^i$ and $S_v^i$; thus, 
they were only exacted to keep the hypotheses rigorously sustained, otherwise the sum of the 
compartments in each population would quickly tend to $N_h$ and $N_v$ in an asymptotic fashion.

Clearly, Figure~\ref{graph_inverse_problem_N} is a reasonable 
description of the outbreak: the general shape of the infection evolution is attained,
all parameters are within realistic possibilities, the $C(t)$ curve overshoots the data by merely $6.00\%$, 
the peak value of $\mathcal{N}_w$ differs from the empirical data maximum by $7.87\%$ and is 
only one week off. However, taking into consideration the order of 
magnitude of the first data point ($C^i = 8{,}201$) and the scale of the infection 
($215{,}319$ probable cases until the $52$th EW \cite{SVS2017}), the TRR output for 
$I_h^i$ ($253{,}360$) is probably too high, even though there is no reference value
to compare with the number of infectious individuals at the beginning of 2016, 
making it difficult to ascertain on a deterministic manner what is a feasible value for $I_h^i$.

\begin{table}[h!]
\vspace{2mm}
\centering
\begin{tabular}{cllllll}
\toprule
$\boldsymbol \alpha$    & TRR input            & $\mathbf{lb}$ & $\mathbf{ub}$ & TRR output & Reference \\
\midrule
$\alpha_{h}$ & $1/5.9$              & $1/12$        & $1/3$      & $1/12$  &
							\cite{Ott2016,Ioos2014,ECDC2015,Valentine2016,Villela2017,Ferguson2016} \\
$\alpha_{v}$ & $1/9.1$              & $1/10$        & $1/5$      & $1/10$  & 
							\cite{Wong2013,Chouin-Carneiro2016,Ferguson2016} \\
$\gamma$     & $1/7.9$              & $1/8.8$       & $1/3$      & $1/8.8$ &
							\cite{Villela2017,Funk2016,Ferguson2016} \\
$\delta$     & $1/11$               & $1/21$        & $1/11$     & $1/16.86$  &  
							\cite{CDCurl,Nelson1986,Villela2017,Freitas2007} \\
$\beta_{h}$  & $1/11.3$             & $1/16.3$      & $1/8$      & $1/16.3$    & 
							\cite{Ferguson2016} \\
$\beta_{v}$  & $1/8.6$              & $1/11.6$      & $1/6.2$    & $1/11.6$    & 
							\cite{Ferguson2016} \\
$S_{h}^{i}$  & $205{,}953{,}959$    & $0.9\times N_h$ & $N_h$    & $205{,}700{,}000$  &
							 ----------- \\
$E_{h}^{i}$  & $8{,}201$            & $0$           & $N_h$      & $15{,}089$ &
							 ----------- \\
$I_{h}^{i}$  & $8{,}201$            & $0$           & $N_h$      & $253{,}360$  &
							 ----------- \\
$S_{v}^{i}$  & $0.99956$             & $0.99$        & $0.999$   & $1$        &
							 ----------- \\
$E_{v}^{i}$  & $2.2 \times 10^{-4}$ & $0$           & $1$        & $0$       &
							 ----------- \\
$I_{v}^{i}$  & $2.2 \times 10^{-4}$ & $0$           & $1$        & $0$     &
							 ----------- \\						 
\bottomrule
\end{tabular}
\caption{TRR setup for the Figure~\ref{graph_inverse_problem_N} calibrated response. The values referring to
parameters are in $\text{days}^{-1}$, human IC are expressed in individuals, and vector quantities in proportion. }
\label{tab_inverse_N}
\end{table}

\begin{figure}[h!]
\centering
\subfigure[Cumulative number]{\includegraphics[scale=0.319]{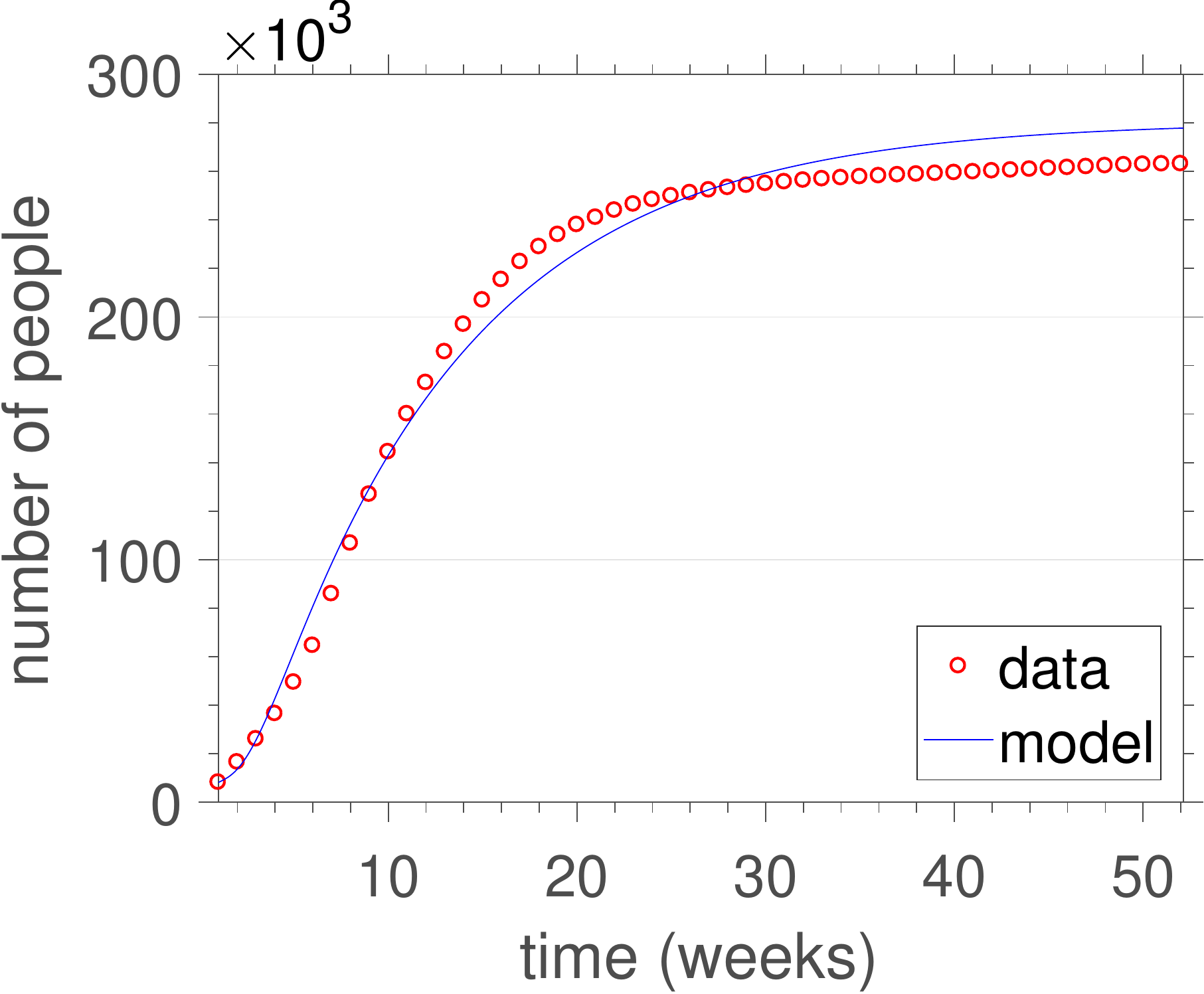}
\label{graph_inverse_C_10k}}
\subfigure[New cases]{\includegraphics[scale=0.32]{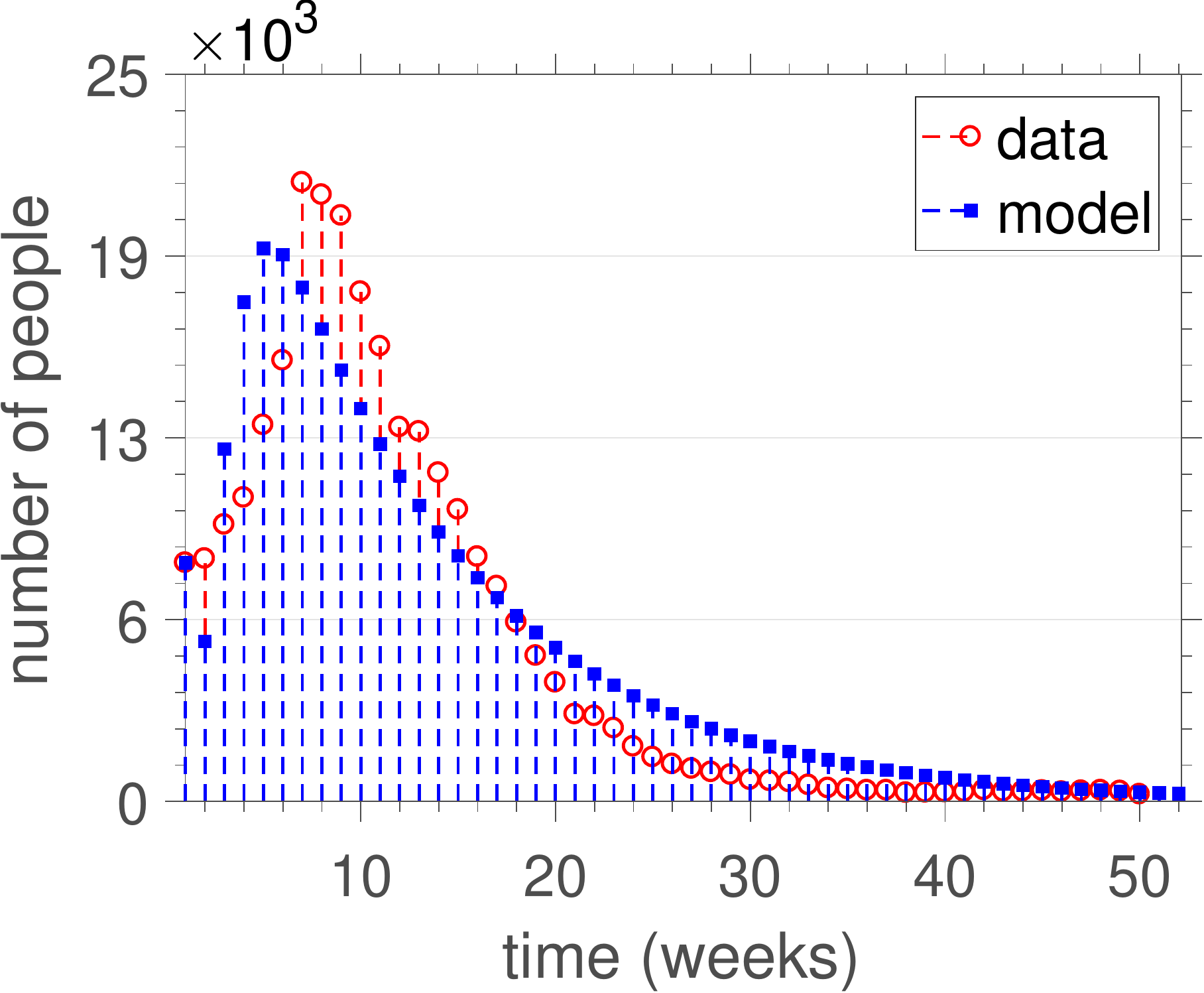}
\label{graph_inverse_NewCases_10k}}\\
\vspace{-3mm}
\caption{Outbreak data (red) and calibrated model response (blue) from Table~\ref{tab_inverse_10k}.}
\label{graph_inverse_problem_10k}
\end{figure}

Figure~\ref{graph_inverse_problem_10k} allows examination of the system behavior when the $I_h^i$ 
value is around $C^i$, by depiction of another result to the inverse problem when the upper 
bound of the initial number of infectious individuals is set to $10{,}000$. The same restriction 
was made over $E_h^i$ merely to simplify the analysis. Table~\ref{tab_inverse_10k} 
displays the resulting parameters and IC.

The model response for a $10{,}000$ $I_h^i$ restriction also presents acceptable 
predictions of the general shape and numbers of the outbreak, even though it is less 
accurate than Figure~\ref{graph_inverse_problem_N} on a fitting criteria for $\mathcal{N}_w$. 
For comparison, the $\mathcal{N}_w$ peak and data maximum difference 
increased to $10.57\%$ and two weeks, while the overshoot on the 
$C(t)$ time series actually reduced to $5.74\%$.

\begin{table}[h!]
\vspace{2mm}
\centering
\begin{tabular}{cllllll}
\toprule
$\boldsymbol \alpha$    & TRR input            & $\mathbf{lb}$ & $\mathbf{ub}$ & TRR output & Reference \\
\midrule
$\alpha_{h}$ & $1/5.9$              & $1/12$        & $1/3$      & $1/12$  &
							\cite{Ott2016,Ioos2014,ECDC2015,Valentine2016,Villela2017,Ferguson2016} \\
$\alpha_{v}$ & $1/9.1$              & $1/10$        & $1/5$      & $1/10$  & 
							\cite{Wong2013,Chouin-Carneiro2016,Ferguson2016} \\
$\gamma$     & $1/7.9$              & $1/8.8$       & $1/3$      & $1/3$ &
							\cite{Villela2017,Funk2016,Ferguson2016} \\
$\delta$     & $1/11$               & $1/21$        & $1/11$     & $1/21$  &  
							\cite{CDCurl,Nelson1986,Villela2017,Freitas2007} \\
$\beta_{h}$  & $1/11.3$             & $1/16.3$      & $1/8$      & $1/10.40$    & 
							\cite{Ferguson2016} \\
$\beta_{v}$  & $1/8.6$              & $1/11.6$      & $1/6.2$    & $1/7.77$    & 
							\cite{Ferguson2016} \\
$S_{h}^{i}$  & $205{,}953{,}959$    & $0.9\times N_h$ & $N_h$ & $205{,}953{,}534$  &
							 ----------- \\
$E_{h}^{i}$  & $8{,}201$            & $0$           & $10{,}000$ & $6{,}827$ &
							 ----------- \\
$I_{h}^{i}$  & $8{,}201$            & $0$           & $10{,}000$ & $10{,}000$  &
							 ----------- \\
$S_{v}^{i}$  & $0.99956$            & $0.99$        & $0.999$    & $0.999586$     &
							 ----------- \\
$E_{v}^{i}$  & $2.2 \times 10^{-4}$ & $0$           & $1$        & $4.14 \times 10^{-4}$       &
							 ----------- \\
$I_{v}^{i}$  & $2.2 \times 10^{-4}$ & $0$           & $1$        & $0$     &
							 ----------- \\
\bottomrule
\end{tabular}
\caption{TRR setup for the Figure~\ref{graph_inverse_problem_10k} calibrated response. The values referring to
parameters are in $\text{days}^{-1}$, human IC are expressed in individuals, and vector quantities in proportion. }
\label{tab_inverse_10k}
\end{table}

\begin{figure}[h!]
\centering
\subfigure[$I_h(t)$ response from Table~\ref{tab_inverse_N}]
{\includegraphics[scale=0.32]{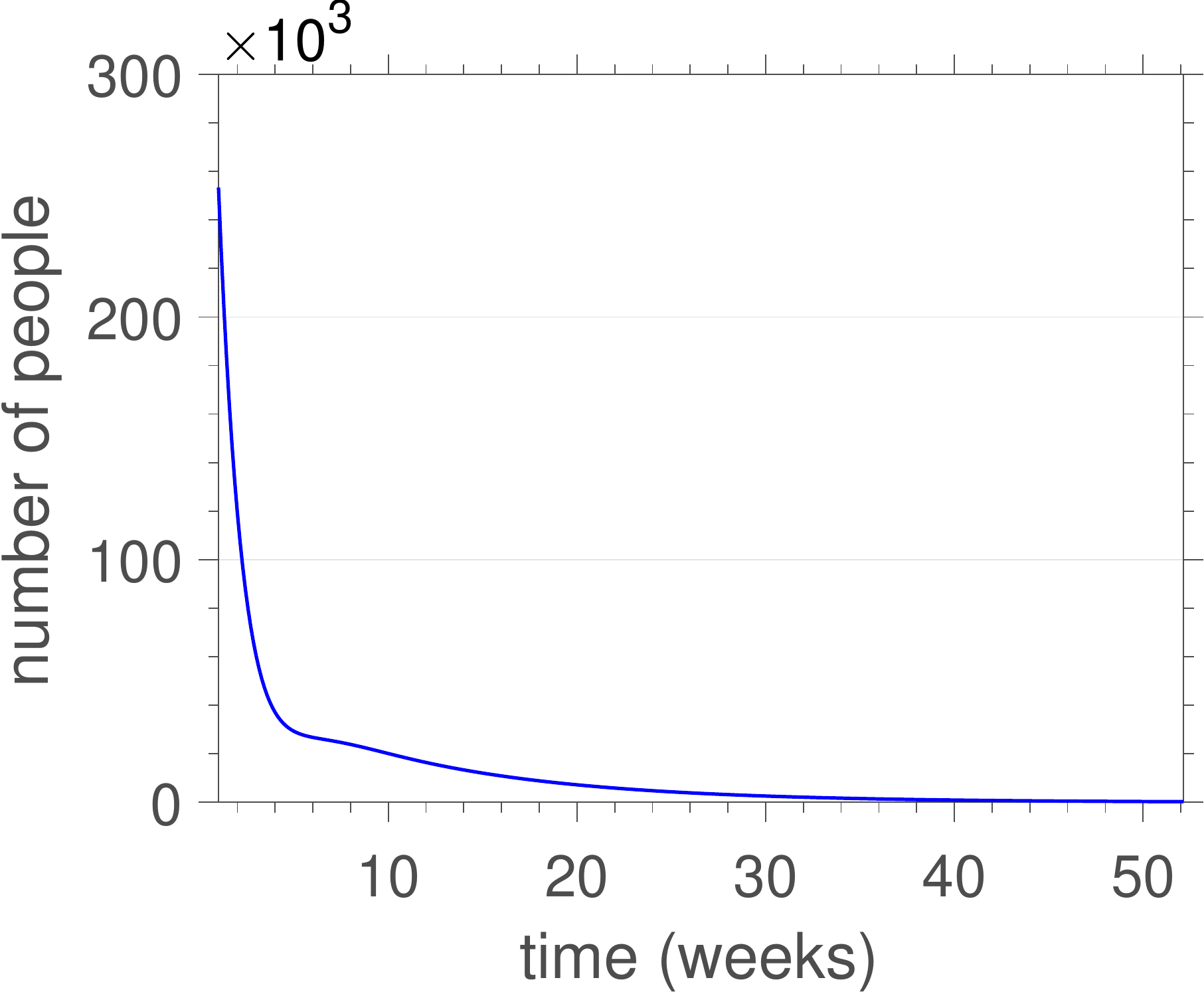} \label{graph_inverse_IH_N}}
\subfigure[$I_h(t)$ response from Table~\ref{tab_inverse_10k}]
{\includegraphics[scale=0.32]{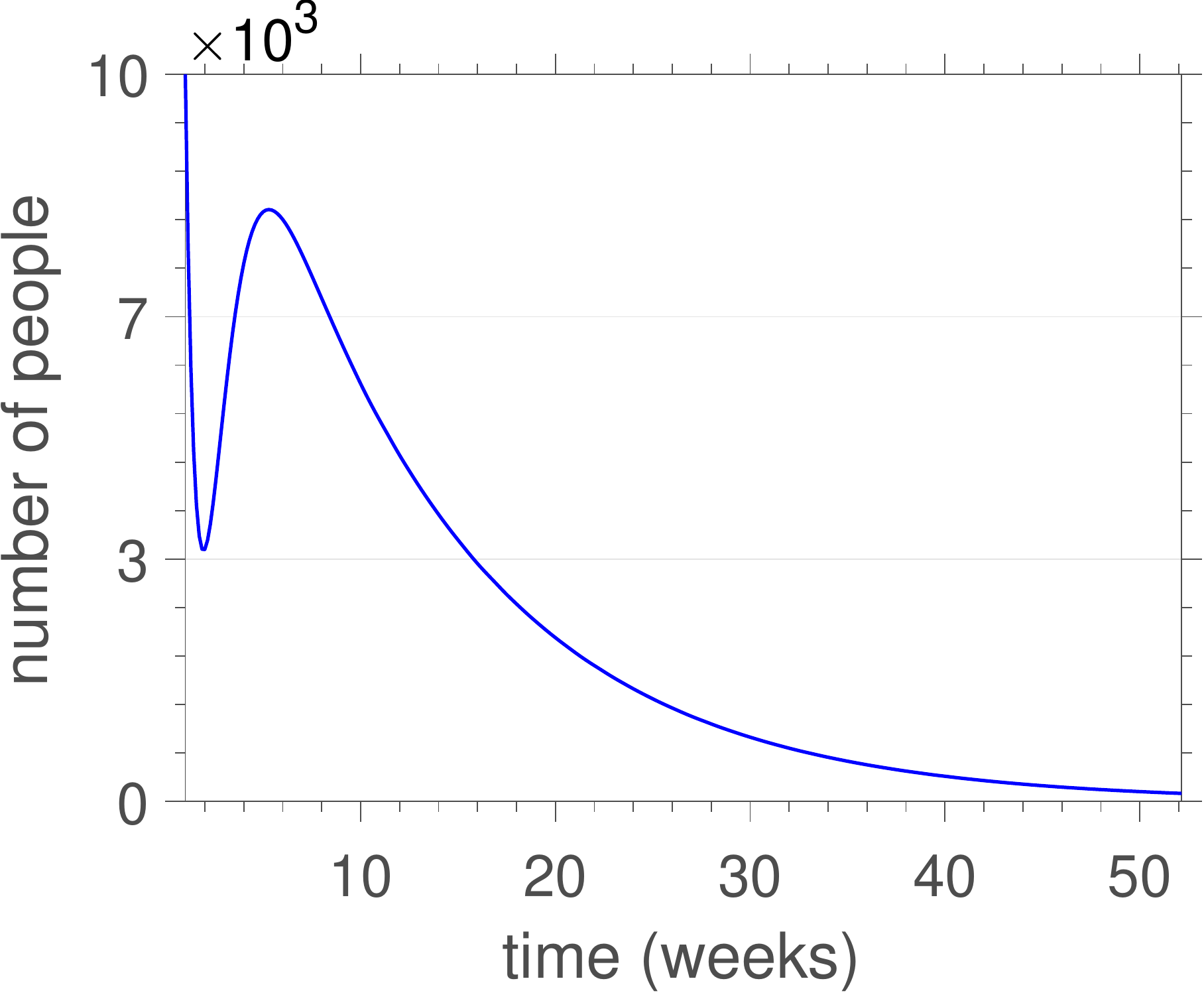}\label{graph_inverse_IH_10k}}\\
\vspace{-3mm}
\caption{Comparison between $I_h(t)$ responses from Table~\ref{tab_inverse_N} (left) and Table~\ref{tab_inverse_10k} (rigth).}
\label{graph_inverse_problem_IH_comparison}
\end{figure}

To compare the two systems defined by the parameters and IC from each table, 
Figure~\ref{graph_inverse_problem_IH_comparison} portrays their $I_h$ response. The system
from Table~\ref{tab_inverse_N} has an almost monotonically decreasing $I_h$ curve, except for a 
slight local maximum around the $6$th EW, while the system from Table~\ref{tab_inverse_10k} 
presents a significant increase in the number of infectious individuals by the same time, which
correspond to the weeks right before the peak infection. As stated, the lack of empirical data
for the current number of infectious at each EW makes it impossible to determine 
what is a quantitatively reasonable prediction for $I_h$ values.
But this work assumes that a more possible scenario involves a 
$I_h$ time series that also follows the general shape of an epidemic curve around the weeks of maximum 
infection, especially considering that Figure~\ref{graph_inverse_IH_N} 
implicates the notion that most people were infected strictly before $2016$, which does not seems
the case suggested by the Brazilian outbreak data of probable cases of infected per EW \cite{SVS2017}
when compared to numbers available for $2015$ \cite{Faria2016}. Thus, with this qualitative criteria 
in mind, the system from Table~\ref{tab_inverse_10k} is selected for a further analysis over its
behavior dependency to the $I_h^i$ initial condition.

\begin{figure}[h!]
\centering
\subfigure[Cumulative number]{\includegraphics[scale=0.32]{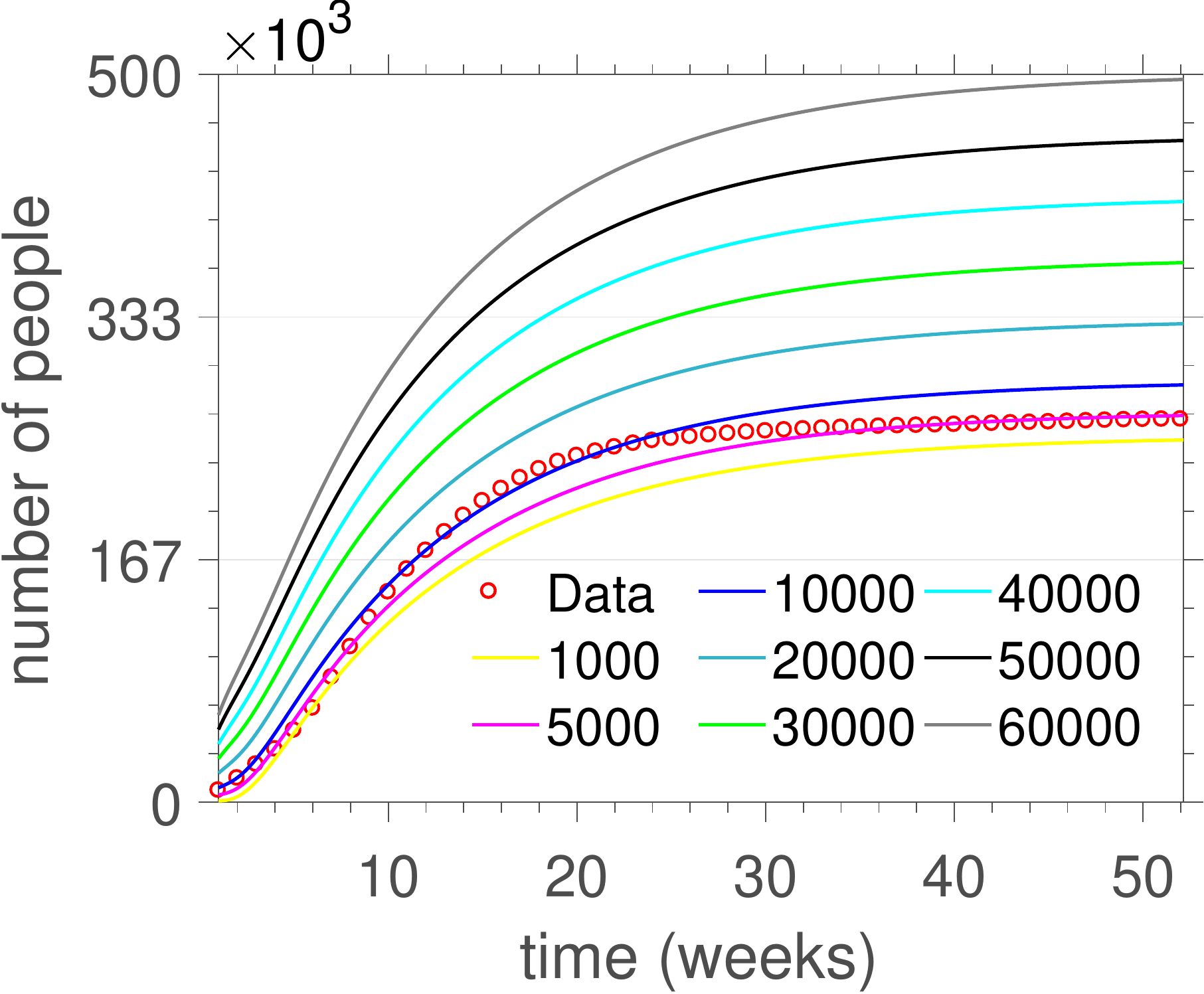} 
\label{graph_inverse_multi_C}}
\subfigure[New cases]{\includegraphics[scale=0.319]{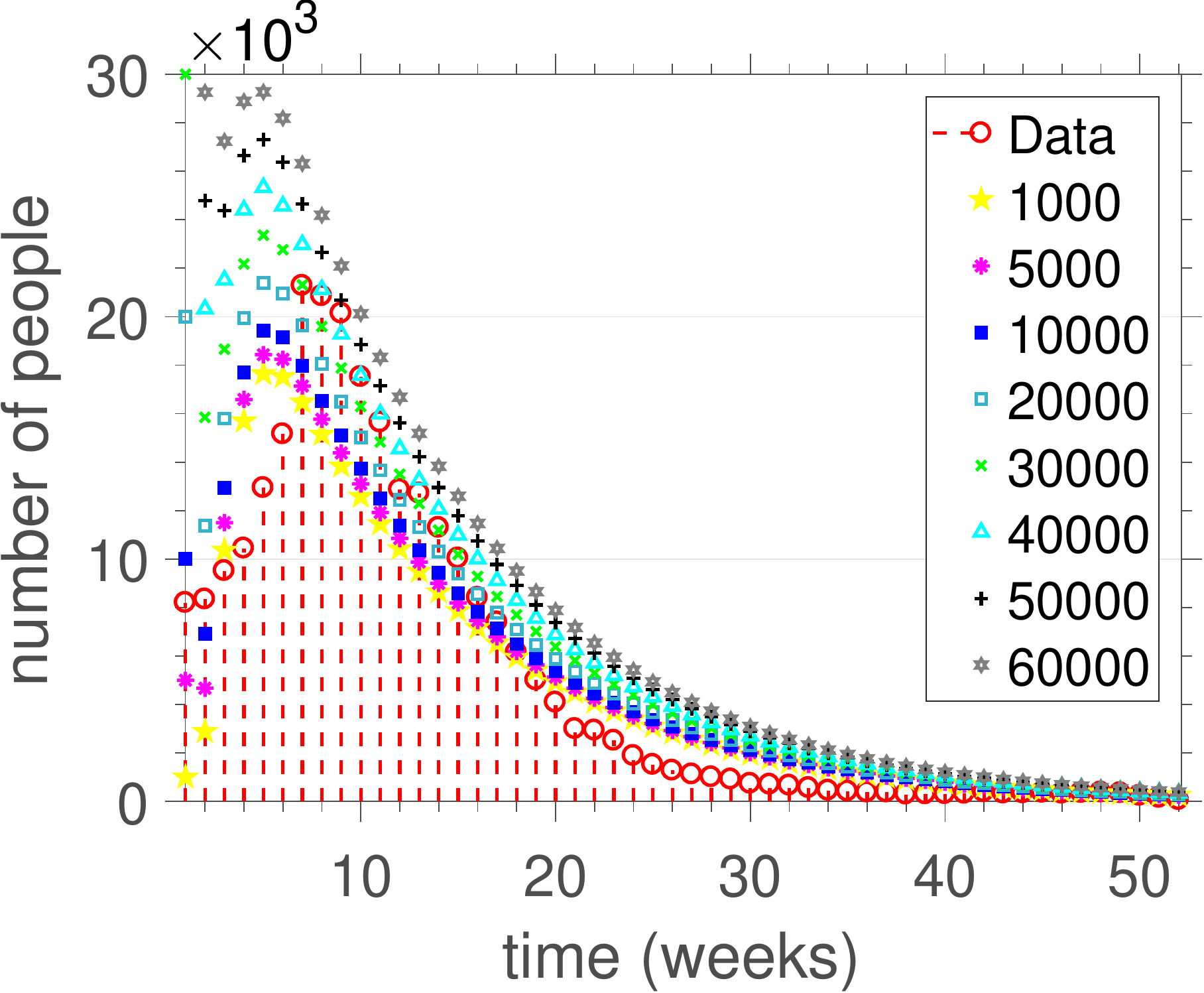}
\label{graph_inverse_multi_NewCases}}\\
\subfigure[$I_h$ response]{\includegraphics[scale=0.32]{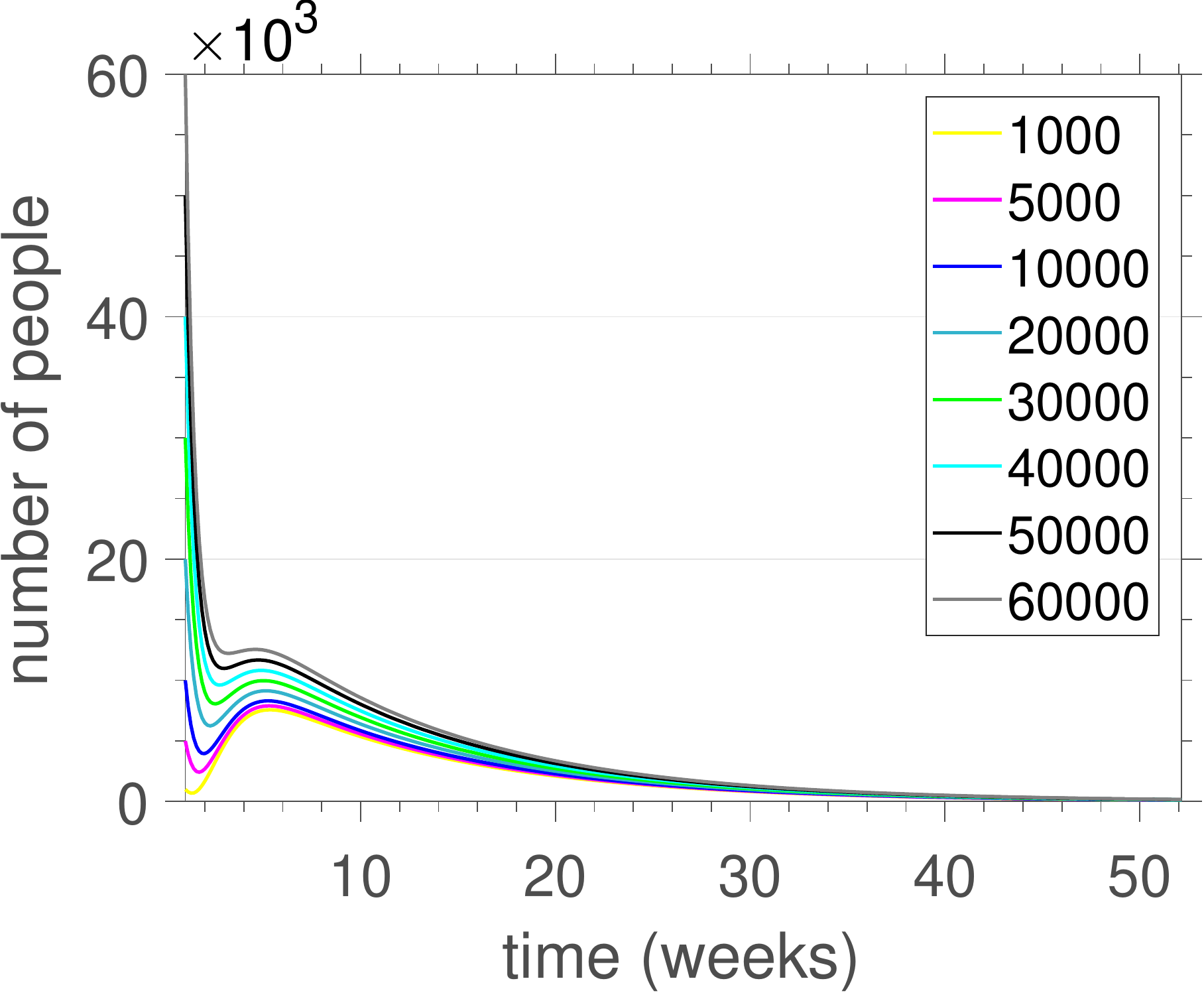} \label{graph_inverse_multi_IH}}
\subfigure[$I_h$ response magnified]{\includegraphics[scale=0.32]{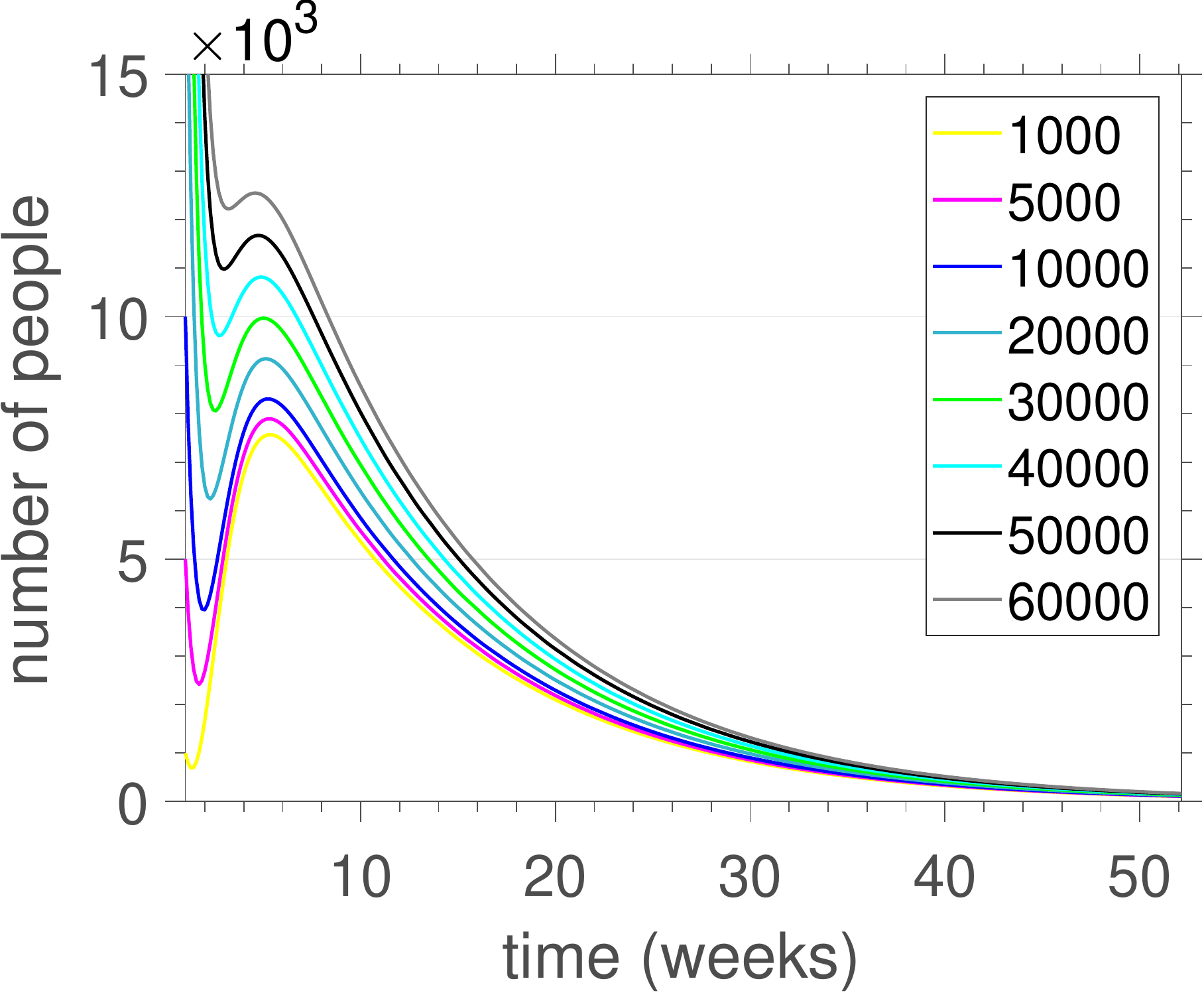}\label{graph_inverse_multi_IH_zoom}}\\
\caption{Multiple $C(t)$, $\mathcal{N}_w$ and $I_h(t)$ responses per EW using the parameters from Table~\ref{tab_inverse_10k}. The values on the legend correspond to the used $I_h^i$ and $E_h^i$ to generate the curves. The other IC are also from Table~\ref{tab_inverse_10k}. The bottom right graph is the $I_h$ comparison magnified around the local maximum region. The red circles are the previously used real data.}
\label{graph_inverse_problem_range}
\end{figure}

Figure~\ref{graph_inverse_problem_range} displays the $C(t)$, $\mathcal{N}_w$ and $I_h(t)$ responses per EW
for various $I_h^i$ on the system with parameters from Table~\ref{tab_inverse_10k}. To simplify the analysis,
$E_h^i$ is considered equal to $I_h^i$ in each case. The remaining IC are the same from the Table.
The pattern suggests that a $I_h^i$ increase on the system with this given set of parameters 
continuously \sanswer{escalates} the $C(t)$ and $\mathcal{N}_w$ curves, eventually overshooting 
the data by far, and \sanswer{reduces} the variations of $I_h(t)$ curve around its local maximum.
Figure~\ref{graph_inverse_problem_range} allows one to make better predictions about the outbreak by
analyzing the multiple possible scenarios of epidemic evolution over different values for the IC missing
empirical information. Clearly, the system response in all cases is qualitative reliable in simulating
the outbreak ($C(t)$ and $\mathcal{N}_w$ shape) and can quantitatively approximate the real data values
for some values of $I_h^i$.

% --------------------------------------------------------------

% --------------------------------------------------------------
\section{Concluding remarks}
\label{concl_remaks}

A SEIR-SEI epidemic model to describe the dynamics of the 2016 Zika virus outbreak in Brazl is developed
and calibrated in this work. Nominal quantities for the parameters are selected from the related literature
concerning the Zika infection, the \emph{Aedes aegypti} genus of mosquitoes, vector-born epidemic models and 
information provided by health organizations. The calibration process is done through the solution of an 
inverse problem with the aid of a Trust-Region-Reflective method, used to pick the best parameter values
that would fit the model response for the number of new infectious cases per week into the disease's
empirical data. Results within realistic values for the parameters are presented, stating reasonable 
descriptions with the curve shape similar to the outbreak evolution and 
proximity between the estimated peak value and data for maximum number of infected during 2016. Further
analysis of the results about the lack of data for an initial condition is performed, exhibiting a range
of values over which the system response keeps its quantitative reliability to a certain degree.

This work is part of a long  project of  modeling and prediction of epidemics related to the Zika virus in the  Brazilian context \cite{Cunhajr_ccis2016,Cunhajr_cnmac2017_1}. In upcoming studies the authors intend
to take into account the uncertainties underlying the model parameters via Bayesian updating 
and employ an Active Subspace approach \cite{Constantine2015,Constantine2018,Constantine2014}
to explore relevant scenarios in parametric studies.
%------------------------------------------------------------------------------

%------------------------------------------------------------------------------
\section*{Acknowledgments}

The authors are indebted to the Brazilian agencies
CNPq (National Council for Scientific and Technological Development), 
CAPES (Coordination for the Improvement of Higher Education Personnel) 
and FAPERJ (Research Support Foundation of the State of Rio de Janeiro)
for the financial support given to this research. They are also grateful to 
Jo\~{a}o Peterson and Vin\'{\i}cius Lopes, both engineering students at UERJ, 
for the collaboration in the initial stages of this work.
The anonymous reviewers made a series of comments 
and suggestions that greatly enriched this paper, for which the authors 
are very grateful.

%------------------------------------------------------------------------------

% References
%------------------------------------------------------------------------------
\section*{References}
%\bibliographystyle{unsrtnat}
%\bibliography{references}

%------------------------------------------------------------------------------

\end{document}

\begin{frontmatter}

\journal{Applied Mathematics and Computation}

\title{Supplementary material B \\[1em] \large Calibration of a SEIR--SEI epidemic model to describe\\ the Zika virus outbreak in Brazil}

\author[uerj]{Eber Dantas}
\ead{eber.paiva@uerj.br}

\author[uerj]{Michel Tosin}
\ead{michel.tosin@uerj.br}

\author[uerj]{Americo Cunha Jr\corref{cor}}
\ead{americo@ime.uerj.br}

\cortext[cor]{Corresponding author.}
%\cortext[a1]{ORCiD: https://orcid.org/0000-0003-2693-0719}
%\cortext[a2]{ORCiD: https://orcid.org/0000-0002-0112-553X}
%\cortext[a3]{ORCiD: https://orcid.org/0000-0002-8342-0363}

\address[uerj]{Universidade do Estado do Rio de Janeiro -- UERJ, \numericoEN{},
			  Rua S\~{a}o Francisco Xavier, 524, Rio de Janeiro - RJ, 20550-900, Brasil.
			  Phone: +55 21 2334-0323 r: 208
}
\end{frontmatter}

% --------------------------------------------------------------
This supplementary material provides working MATLAB scripts for the main document 
``Calibration of a SEIR-SEI epidemic model to describe the Zika virus outbreak in Brazil.'' The first 
section uses ``ode45'' function to evaluate the forward problem as explained in section 3.1 of the main
document. Section 2 solves the inverse problem via a TRR algorithm (main document section 3.2) implemented in 
the ``lsqcurvefit'' function of the software. Finally, section 3 exemplifies a variation of the model by
introducing vital dynamics for the human population, useful for comparison.

All the codes in this document use the following data set. See supplementary material A.

\begin{lstlisting}[style=Matlab-editor,basicstyle=\footnotesize,caption = Data set]
% New Cases of infection per epidemiological week [51]
NewCasesData = [8201,8344,9516,10443,12940,15158,21280, ...
20854,20139,17520,15644,12865,12715,11298,10034,8410,   ...
7397,6154,5008,4089,2997,2936,2516,1888,1521,1282,1126, ...
1010,917,741,703,665,564,455,422,380,380,296,305,315,   ...
329,390,338,346,370,331,361,381,353,244,159,69];

% Cumulative cases
CData = cumsum(NewCasesData);

% Total number of cases in 2015 [13]
C2015 = 29639;
\end{lstlisting}

\section{Forward Problem}

The default values for the parameters and initial conditions (IC) are the TRR output from Table 2, 
section 3.3 of the main document.

\begin{lstlisting}[style=Matlab-editor,basicstyle=\footnotesize,caption = Forward problem - main file]
% Parameters
Nh = 206e6;		% Human population
bH = 1/10.40;		% Vector-to-Human transmission rate
aH = 1/12;		% Intrinsic incubation rate
yH = 1/3;		% Infectious rate	

Nv = 1;			% Vector population
bV = 1/7.77;		% Human-to-Vector transmission rate
aV = 1/10;		% Extrinsic incubation rate
dV = 1/21;		% Inverse of vector lifespan

param = [Nh bH aH yH Nv bV aV dV];

% Initial Conditions
SH0 = 205953534;	% Susceptible humans
EH0 = 6827;		% Exposed humans
IH0 = 10000;		% Infectious humans
RH0 = C2015;		% Recovered humans
SV0 = 0.999586;		% Susceptible vectors
EV0 = 4.14e-4;		% Exposed vectors
IV0 = 0;		% Infectious vectors
C0  = NewCasesData(1);	% Cumulative number

IC = [SH0 EH0 IH0 RH0 SV0 EV0 IV0 C0];

% Integrate the IVP
dt = 0.1;		% Time step (days)
tspan = 7:dt:365; 	% Domain discretization

opt = odeset('RelTol',1.0e-6,'AbsTol',1.0e-9);
[time y] = ode45(@(t,y)rhs_SEIR_SEI(t,y,param),tspan,IC,opt);

% Model response
SH = y(:,1); EH = y(:,2); IH = y(:,3); RH = y(:,4);
SV = y(:,5); EV = y(:,6); IV = y(:,7);

C  = y(:,8);

NewCases = C(1);
for i = 2:52
    NewCases(i) = C( (7*i-7)/dt +1 ) - C( (7*(i-1)-7)/dt +1 );
end

% Plotting
   %Week timescale: time/7

figure(1)
plot(time/7,C)
hold on
scatter([1:52],CData)

figure(2)
stem(NewCasesData)
hold on
stem(NewCases)
\end{lstlisting}

\begin{lstlisting}[style=Matlab-editor,basicstyle=\footnotesize,caption = Forward problem - right hand side]
function ydot = rhs_SEIR_SEI(t,y,param)

% phys_param = [Nh bH aH yH Nv bV aV dV];
Nh  = param(1); bH  = param(2); aH  = param(3); yH  = param(4);
Nv  = param(5); bV  = param(6); aV  = param(7); dV  = param(8);

% System of ODEs 

% y = [SH EH IH RH SV EV IV C] is the solution vector
ydot = zeros(size(y));
ydot(1) = - bH*y(1).*y(7)/Nv;       		% dSh/dt
ydot(2) = bH*y(1).*y(7)/Nv - aH*y(2);      	% dEh/dt
ydot(3) = aH*y(2) - yH*y(3);			% dIh/dt
ydot(4) = yH*y(3);				% dRh/dt
ydot(5) = dV - bV.*y(5).*y(3)/Nh - dV*y(5);    	% dSv/dt
ydot(6) = bV*y(5).*y(3)/Nh - aV*y(6) - dV*y(6);	% dEv/dt
ydot(7) = aV*y(6) - dV*y(7);			% dIv/dt
ydot(8) = aH*y(2);				% dC/dt

end                 
\end{lstlisting}

\section{Inverse Problem}

See section 2.3 and 2.4 of the main document for the nominal parameters and initial conditions used as
initial guesses here. The default results are those from Table 2, section 3.3 of the main document. 

\begin{lstlisting}[style=Matlab-editor,basicstyle=\footnotesize,
caption = Inverse problem main file]
% Nominal Parameters
Nh = 206e6;		% Human population
bH = 1/11.3;		% Vector-to-Human transmission rate
aH = 1/5.9;		% Intrinsic incubation rate
yH = 1/7.9;		% Infectious rate	

Nv = 1;			% Vector population
bV = 1/8.6;		% Human-to-Vector transmission rate
aV = 1/9.1;		% Extrinsic incubation rate
dV = 1/11;		% Inverse of vector lifespan  

% Nominal Initial Conditions
IH0 = NewCasesData(1);		% Infectious humans
EH0 = IH0;			% Exposed humans
RH0 = C2015;			% Recovered humans
SH0 = Nh - (IH0 + EH0 + RH0);	% Susceptible humans

IV0 = 0.00022;		% Infectious vectors
EV0 = IV0;		% Exposed vectors
SV0 = Nv - (EV0 + IV0);	% Susceptible vectors

C0  = NewCasesData(1);	% Cumulative number

% TRR initial guess (variable_TRRparam)
	% RH0 is omitted to optimize the procedure
inital_guess = [bH aH yH bV aV dV SH0 EH0 IH0 SV0 EV0 IV0];
fixed_TRRparam = [Nh Nv C0]; 

% Data for TRR
data = [NewCasesData Nh-RH0 Nv]; % Last two values are for the 
				 % sum comparisons
data_pos = [7:7:365 400 450]; 	 % Position for comparison between 
				 % data and NewCases. Last 2 values 
				 % are arbitrary.

% Lower bound of variable parameters
lb = [1/16.3, 1/12, 1/8.8, 1/11.6, 1/10, 1/21, 0.9*Nh, ...
	0,     0,  0.99,  0,  0];
% upper bound of variable parameters
up = [   1/8,  1/3,  1/3,   1/6.2,  1/5,  1/11,    Nh, ... 
    10000, 10000, 0.999, Nv, Nv];                

% Time discretization
dt = 1;		% Time step (days)
tspan = 7:dt:365;

% Output Function
F = @(variable_TRRparam,data_pos)TRR_FunctionOutput(...
	variable_TRRparam,data_pos,fixed_TRRparam,tspan,dt); 

% Algorithm options
options = optimoptions('lsqcurvefit','Display','iter', ... 
	  'Algorithm','trust-region-reflective', ... 
	  'MaxIterations',100000,'MaxFunctionEvaluations',  ... 
	  500000,'FunctionTolerance',1e-007,'StepTolerance', ...
	  1e-007,'FiniteDifferenceType','central');
 
% TRR procedure
result_TRRparam = lsqcurvefit(F,inital_guess,data_pos,data,lb, ...
		  up,options);

% New set of parameters and IC
new_param = [Nh result_TRRparam(1:3) Nv result_TRRparam(4:6)];
new_IC = [result_TRRparam(7:9) RH0 result_TRRparam(10) ...
	  result_TRRparam(11:12) C0];
	  
% Compartmentalization hypotheses correction
Xh = Nh - RH0 - sum(result_TRRparam(7:9));
Xv = Nv - sum(result_TRRparam(10:12));
new_IC(1) = result_TRRparam(7) + Xh;
new_IC(5) = result_TRRparam(10) + Xv;

% Integrate the new IVP
[time y] = ode45(@(t,y)rhs_SEIR_SEI(t,y,new_param),tspan,new_IC);
opt = odeset('RelTol',1.0e-6,'AbsTol',1.0e-9);	% ode45 options

% Model response
SH = y(:,1); EH = y(:,2); IH = y(:,3); RH = y(:,4);
SV = y(:,5); EV = y(:,6); IV = y(:,7);

C  = y(:,8);

NewCases = C(1);
for i = 2:52
    NewCases(i) = C( (7*i-7)/dt +1 ) - C( (7*(i-1)-7)/dt +1 );
end

%Plotting
	%week timescale: time/7
figure(1)
plot(time/7,C)
hold on
scatter([1:52],CData)

figure(2)
stem(NewCasesData)
hold on
stem(NewCases)
\end{lstlisting}

\begin{lstlisting}[style=Matlab-editor,basicstyle=\footnotesize,
caption = Inverse problem output function]
function F = TRR_FunctionOutput(variable_TRRparam,data_pos,fixed_TRRparam,tspan,dt)

% Adjusting IC vector
 % fixed_TRRparam = [Nh Nv C0]; 
 % variable_TRRparam = [bH aH yH bV aV dV SH0 EH0 IH0 SV0 EV0 IV0]	
 % IC = [SH0 EH0 IH0 SV0 EV0 IV0 C0]
IC = [variable_TRRparam(7:12) fixed_TRRparam(3)];

% Integrate the IVP for function evaluation
 % y = [SH EH IH SV EV IV C] is the solution vector
 % RH is omitted to optimize the procedure
opt = odeset('RelTol',1.0e-6,'AbsTol',1.0e-9);
[time y] = ode45(@(t,y)TRR_rhs_SEIR_SEI(t,y,variable_TRRparam, ...
		fixed_TRRparam),tspan,IC,opt);

% SumH = SH0+EH0+IH0 (is compared to Nh-R0)
SumH = y(1,1) + y(1,2) + y(1,3);  

% SumV = SV0+EV0+IV0 (is compared to Nv)
SumV = y(1,4) + y(1,5) + y(1,6);  

C  = y(:,7);

NewCases = C(1);
for i = 2:52
    NewCases(i) = C( (7*i-7)/dt +1 ) - C( (7*(i-1)-7)/dt +1 );
end

% Output value
F = [NewCases SumH SumV];

end
\end{lstlisting}

\begin{lstlisting}[style=Matlab-editor,basicstyle=\footnotesize,
caption = Inverse problem right hand side]
function ydot = TRR_rhs_SEIR_SEI(t,y,variable_TRRparam, ...
				fixed_TRRparam)
% fixed_TRRparam = [N C0];
% variable_TRRparam = [bH aH yH bV aV dV SH0 EH0 IH0 SV0 EV0 IV0];

Nh   = fixed_TRRparam(1);    % human population size 
bH  = variable_TRRparam(1);  % vector-to-human transmission rate     
aH  = variable_TRRparam(2);  % incubation human frequency 
yH  = variable_TRRparam(3);  % infectious human frequency 

Nv = fixed_TRRparam(2);
bV  = variable_TRRparam(4);  % human-to-vector transmission rate 
aV  = variable_TRRparam(5);  % incubation vector frequency
dV  = variable_TRRparam(6);  % vector lifespan frequency

% System of ODEs 
 % y = [SH EH IH SV EV IV C] is the solution vector
 % RH is omitted to optimize the procedure
    
ydot = zeros(size(y));


ydot(1) = -bH*y(1).*y(6)/Nv;			  % dSh/dt
ydot(2) = bH*y(1).*y(6)/Nv - aH*y(2);		  % dEh/dt
ydot(3) = aH*y(2) - yH*y(3);			  % dIh/dt
ydot(4) = dV - bV.*y(4).*(y(3)/Nh) - dV*y(4);	  % dSv/dt
ydot(5) = bV*y(4).*(y(3)/Nh) - aV*y(5) - dV*y(5); % dEv/dt
ydot(6) = aV*y(5) - dV*y(6);			  % dIv/dt
ydot(7) = aH*y(2);				  % dC/dt

end
\end{lstlisting}

\section{Changing human population}

A recruitment term ($\eta\,N_h$) and a mortality rate ($\mu$) are needed to implement a 
changing human population. The human variables in Eqs.~(1) in section 2.2 of the main document become

\begin{equation}
\begin{aligned}
\dod{S_h}{t} &= \eta\,N_h -\beta_h \, {S_h} \, \frac{I_v}{N_v} - \mu\,S_h \:,\\ 
\dod{E_h}{t} &= \beta_h \, S_h \, \frac{I_v}{N_v} - \alpha_h \, E_h - \mu\,E_h \:,\\
\dod{I_h}{t} &= \alpha_h \, E_h - \gamma \, I_h - \mu\,I_h \:,\\
\dod{R_h}{t} &= \gamma \, I_h  - \mu\,R_h \:.
\end{aligned}
\end{equation}

\noindent
Also, a new differential equation must be added to the model to track the varying population

\begin{equation}
\dod{N_h}{t} = \eta\,N_h - \mu\,(S_h+E_h+I_h+R_h) \:,
\end{equation}

\noindent
and, consequently, its respective initial condition, $N_h(0)$, is required for the forward problem.
Code~\ref{vital} provides the right hand side of this variation of the model.

\begin{lstlisting}[style=Matlab-editor,basicstyle=\footnotesize,
caption = Vital Dynamics - rigth hand side,label = vital]
function ydot = rhs_SEIR_SEI(t,y,phys_param)

% param = [nH bH aH yH bV aV dV mH];

nH  = phys_param(1);    % human birth rate (per person) per day
bH  = phys_param(2);    % vector-to-human transmission rate
aH  = phys_param(3);    % incubation human frequency
yH  = phys_param(4);    % infectious human frequency
bV  = phys_param(5);    % human-to-vector transmission rate
aV  = phys_param(6);    % incubation vector frequency
dV  = phys_param(7);    % vector lifespan frequency 
mH  = phys_param(8);    % human death rate (per person) per day


% System of ODEs 
% y = [SH EH IH RH SV EV IV C NH] is the solution vector

ydot = zeros(size(y));


ydot(1) = nH*y(9) - bH*y(1).*y(7) - mH*y(1);		% dSh/dt
ydot(2) = bH*y(1).*y(7) - aH*y(2) - mH*y(2);		% dEh/dt
ydot(3) = aH*y(2) - yH*y(3) - mH*y(3);			% dIh/dt
ydot(4) = yH*y(3) - mH*y(4);				% dRh/dt
ydot(5) = dV - bV.*y(5).*(y(3)/y(9)) - dV*y(5);		% dSv/dt
ydot(6) = bV*y(5).*(y(3)/y(9)) - aV*y(6) - dV*y(6);	% dEv/dt
ydot(7) = aV*y(6) - dV*y(7);				% dIv/dt
ydot(8) = aH*y(2);					% dC/dt
ydot(9) = nH*y(9) - mH*(y(1)+y(2)+y(3)+y(4));		% dNh/dt
                                                                        
end
\end{lstlisting}

An estimation for the values of $\eta$ and $\mu$ (per thousand individuals in a year) can be found in 
\cite{IBGE} for 2015 (the most recent officially published mortality and birth rates for Brazil).
The solution of the forward problem with this code can be used to observe how the human vital dynamics do 
not affect $\mathcal{N}_{w}$ (neither $C(t)$) in a one year outbreak for this model. Of course, $S_h(t)$ 
will behave different, as expected.